\documentstyle[a4,12pt]{article}
\begin{document}
\def\seCtion#1{\section{#1} \setcounter{equation}{0}}
\renewcommand\theequation{\ifnum\value{section}>0%
  {\thesection.\arabic{equation}}\fi}
\def\temp{1.35}%
\let\tempp=\relax
\expandafter\ifx\csname psboxversion\endcsname\relax
  \message{PSBOX(\temp)}%
\else
    \ifdim\temp cm>\psboxversion cm
      \message{PSBOX(\temp)}%
    \else
      \message{PSBOX(\psboxversion) is already loaded: I won't load
        PSBOX(\temp)!}%
      \let\temp=\psboxversion
      \let\tempp= 
    \fi
\fi
\tempp
\message{by Jean Orloff: loading ...}
\let\psboxversion=\temp
\catcode`\@=11
%
%
\def\psfortextures{
\def\PSspeci@l##1##2{%
\special{illustration ##1\space scaled ##2}%
}}%
\def\psfordvitops{
\def\PSspeci@l##1##2{%
\special{dvitops: import ##1\space \the\drawingwd \the\drawinght}%
}}%
\def\psfordvips{
\def\PSspeci@l##1##2{%
\d@my=0.1bp \d@mx=\drawingwd \divide\d@mx by\d@my
\includegraphics{##1\space}}}%
\def\psforoztex{
\def\PSspeci@l##1##2{%
\special{##1 \space
      ##2 1000 div dup scale
      \number-\psllx\space\space \number-\pslly\space\space translate
}}}%
\def\psfordvitps{
\def\dvitpsLiter@ldim##1{\dimen0=##1\relax
\special{dvitps: Literal "\number\dimen0\space"}}%
\def\PSspeci@l##1##2{%
\at(0bp;\drawinght){%
\special{dvitps: Include0 "psfig.psr"}
\dvitpsLiter@ldim{\drawingwd}%
\dvitpsLiter@ldim{\drawinght}%
\dvitpsLiter@ldim{\psllx bp}%
\dvitpsLiter@ldim{\pslly bp}%
\dvitpsLiter@ldim{\psurx bp}%
\dvitpsLiter@ldim{\psury bp}%
\special{dvitps: Literal "startTexFig"}%
\special{dvitps: Include1 "##1"}%
\special{dvitps: Literal "endTexFig"}%
}}}%
\def\psfordvialw{
\def\PSspeci@l##1##2{
\special{language "PostScript",
position = "bottom left",
literal "  \psllx\space \pslly\space translate
  ##2 1000 div dup scale
  -\psllx\space -\pslly\space translate",
include "##1"}
}}%
\def\psforptips{
\def\PSspeci@l##1##2{{
\d@mx=\psurx bp
\advance \d@mx by -\psllx bp
\divide \d@mx by 1000\multiply\d@mx by \xscale
\incm{\d@mx}
\let\tmpx\dimincm
\d@my=\psury bp
\advance \d@my by -\pslly bp
\divide \d@my by 1000\multiply\d@my by \xscale
\incm{\d@my}
\let\tmpy\dimincm
\d@mx=-\psllx bp
\divide \d@mx by 1000\multiply\d@mx by \xscale
\d@my=-\pslly bp
\divide \d@my by 1000\multiply\d@my by \xscale
\at(\d@mx;\d@my){\special{ps:##1 x=\tmpx cm, y=\tmpy cm}}
}}}%
\def\psonlyboxes{
\def\PSspeci@l##1##2{%
\at(0cm;0cm){\boxit{\vbox to\drawinght
  {\vss\hbox to\drawingwd{\at(0cm;0cm){\hbox{({\tt##1})}}\hss}}}}
}}%
\def\psloc@lerr#1{%
\let\savedPSspeci@l=\PSspeci@l%
\def\PSspeci@l##1##2{%
\at(0cm;0cm){\boxit{\vbox to\drawinght
  {\vss\hbox to\drawingwd{\at(0cm;0cm){\hbox{({\tt##1}) #1}}\hss}}}}
\let\PSspeci@l=\savedPSspeci@l
}}%
%
%
\newread\pst@mpin
\newdimen\drawinght\newdimen\drawingwd
\newdimen\psxoffset\newdimen\psyoffset
\newbox\drawingBox
\newcount\xscale \newcount\yscale \newdimen\pscm\pscm=1cm
\newdimen\d@mx \newdimen\d@my
\newdimen\pswdincr \newdimen\pshtincr
\let\ps@nnotation=\relax
{\catcode`\|=0 |catcode`|\=12 |catcode`|
|catcode`#=12 |catcode`*=14
|xdef|backslashother{\}*
|xdef|percentother{
|xdef|tildeother{~}*
|xdef|sharpother{#}*
}%
\def\R@moveMeaningHeader#1:->{}%
\def\uncatcode#1{%
\edef#1{\expandafter\R@moveMeaningHeader\meaning#1}}%
\def\execute#1{#1}
\def\psm@keother#1{\catcode`#112\relax}
\def\executeinspecs#1{%
\execute{\begingroup\let\do\psm@keother\dospecials\catcode`\^^M=9#1\endgroup}}%
\def\@mpty{}%
\def\matchexpin#1#2{
  \fi%
  \edef\tmpb{{#2}}%
  \expandafter\makem@tchtmp\tmpb%
  \edef\tmpa{#1}\edef\tmpb{#2}%
  \expandafter\expandafter\expandafter\m@tchtmp\expandafter\tmpa\tmpb\endm@tch%
  \if\match%
}%
\def\matchin#1#2{%
  \fi%
  \makem@tchtmp{#2}%
  \m@tchtmp#1#2\endm@tch%
  \if\match%
}%
\def\makem@tchtmp#1{\def\m@tchtmp##1#1##2\endm@tch{%
  \def\tmpa{##1}\def\tmpb{##2}\let\m@tchtmp=\relax%
  \ifx\tmpb\@mpty\def\match{YN}%
  \else\def\match{YY}\fi%
}}%
\def\incm#1{{\psxoffset=1cm\d@my=#1
 \d@mx=\d@my
  \divide\d@mx by \psxoffset
  \xdef\dimincm{\number\d@mx.}
  \advance\d@my by -\number\d@mx cm
  \multiply\d@my by 100
 \d@mx=\d@my
  \divide\d@mx by \psxoffset
  \edef\dimincm{\dimincm\number\d@mx}
  \advance\d@my by -\number\d@mx cm
  \multiply\d@my by 100
 \d@mx=\d@my
  \divide\d@mx by \psxoffset
  \xdef\dimincm{\dimincm\number\d@mx}
}}%
%
\newif\ifNotB@undingBox
\newhelp\PShelp{Proceed: you'll have a 5cm square blank box instead of
your graphics.}%
\def\s@tsize#1 #2 #3 #4\@ndsize{
  \def\psllx{#1}\def\pslly{#2}%
  \def\psurx{#3}\def\psury{#4}
  \ifx\psurx\@mpty\NotB@undingBoxtrue
  \else
    \drawinght=#4bp\advance\drawinght by-#2bp
    \drawingwd=#3bp\advance\drawingwd by-#1bp
  \fi
  }%
\def\sc@nBBline#1:#2\@ndBBline{\edef\p@rameter{#1}\edef\v@lue{#2}}%
\def\g@bblefirstblank#1#2:{\ifx#1 \else#1\fi#2}%
{\catcode`\%=12
\xdef\B@undingBox{
\def\ReadPSize#1{
 \readfilename#1\relax
 \let\PSfilename=\lastreadfilename
 \openin\pst@mpin=#1\relax
 \ifeof\pst@mpin \errhelp=\PShelp
   \errmessage{I haven't found your postscript file (\PSfilename)}%
   \psloc@lerr{was not found}%
   \s@tsize 0 0 142 142\@ndsize
   \closein\pst@mpin
 \else
   \if\matchexpin{\GlobalInputList}{, \lastreadfilename}%
   \else\xdef\GlobalInputList{\GlobalInputList, \lastreadfilename}%
     \immediate\write\psbj@inaux{\lastreadfilename,}%
   \fi%
   \loop
     \executeinspecs{\catcode`\ =10\global\read\pst@mpin to\n@xtline}%
     \ifeof\pst@mpin
       \errhelp=\PShelp
       \errmessage{(\PSfilename) is not an Encapsulated PostScript File:
           I could not find any \B@undingBox: line.}%
       \edef\v@lue{0 0 142 142:}%
       \psloc@lerr{is not an EPSFile}%
       \NotB@undingBoxfalse
     \else
       \expandafter\sc@nBBline\n@xtline:\@ndBBline
       \ifx\p@rameter\B@undingBox\NotB@undingBoxfalse
         \edef\t@mp{%
           \expandafter\g@bblefirstblank\v@lue\space\space\space}%
         \expandafter\s@tsize\t@mp\@ndsize
       \else\NotB@undingBoxtrue
       \fi
     \fi
   \ifNotB@undingBox\repeat
   \closein\pst@mpin
 \fi
\message{#1}%
}%
%
%
\def\psboxto(#1;#2)#3{\vbox{%
   \ReadPSize{#3}%
   \advance\pswdincr by \drawingwd
   \advance\pshtincr by \drawinght
   \divide\pswdincr by 1000
   \divide\pshtincr by 1000
   \d@mx=#1
   \ifdim\d@mx=0pt\xscale=1000
         \else \xscale=\d@mx \divide \xscale by \pswdincr\fi
   \d@my=#2
   \ifdim\d@my=0pt\yscale=1000
         \else \yscale=\d@my \divide \yscale by \pshtincr\fi
   \ifnum\yscale=1000
         \else\ifnum\xscale=1000\xscale=\yscale
                    \else\ifnum\yscale<\xscale\xscale=\yscale\fi
              \fi
   \fi
   \divide\drawingwd by1000 \multiply\drawingwd by\xscale
   \divide\drawinght by1000 \multiply\drawinght by\xscale
   \divide\psxoffset by1000 \multiply\psxoffset by\xscale
   \divide\psyoffset by1000 \multiply\psyoffset by\xscale
   \global\divide\pscm by 1000
   \global\multiply\pscm by\xscale
   \multiply\pswdincr by\xscale \multiply\pshtincr by\xscale
   \ifdim\d@mx=0pt\d@mx=\pswdincr\fi
   \ifdim\d@my=0pt\d@my=\pshtincr\fi
   \message{scaled \the\xscale}%
 \hbox to\d@mx{\hss\vbox to\d@my{\vss
   \global\setbox\drawingBox=\hbox to 0pt{\kern\psxoffset\vbox to 0pt{%
      \kern-\psyoffset
      \PSspeci@l{\PSfilename}{\the\xscale}%
      \vss}\hss\ps@nnotation}%
   \global\wd\drawingBox=\the\pswdincr
   \global\ht\drawingBox=\the\pshtincr
   \global\drawingwd=\pswdincr
   \global\drawinght=\pshtincr
   \baselineskip=0pt
   \copy\drawingBox
 \vss}\hss}%
  \global\psxoffset=0pt
  \global\psyoffset=0pt
  \global\pswdincr=0pt
  \global\pshtincr=0pt 
  \global\pscm=1cm 
}}%
%
%
\def\psboxscaled#1#2{\vbox{%
  \ReadPSize{#2}%
  \xscale=#1
  \message{scaled \the\xscale}%
  \divide\pswdincr by 1000 \multiply\pswdincr by \xscale
  \divide\pshtincr by 1000 \multiply\pshtincr by \xscale
  \divide\psxoffset by1000 \multiply\psxoffset by\xscale
  \divide\psyoffset by1000 \multiply\psyoffset by\xscale
  \divide\drawingwd by1000 \multiply\drawingwd by\xscale
  \divide\drawinght by1000 \multiply\drawinght by\xscale
  \global\divide\pscm by 1000
  \global\multiply\pscm by\xscale
  \global\setbox\drawingBox=\hbox to 0pt{\kern\psxoffset\vbox to 0pt{%
     \kern-\psyoffset
     \PSspeci@l{\PSfilename}{\the\xscale}%
     \vss}\hss\ps@nnotation}%
  \advance\pswdincr by \drawingwd
  \advance\pshtincr by \drawinght
  \global\wd\drawingBox=\the\pswdincr
  \global\ht\drawingBox=\the\pshtincr
  \global\drawingwd=\pswdincr
  \global\drawinght=\pshtincr
  \baselineskip=0pt
  \copy\drawingBox
  \global\psxoffset=0pt
  \global\psyoffset=0pt
  \global\pswdincr=0pt
  \global\pshtincr=0pt 
  \global\pscm=1cm
}}%
%
\def\psbox#1{\psboxscaled{1000}{#1}}%
\newif\ifn@teof\n@teoftrue
\newif\ifc@ntrolline
\newif\ifmatch
\newread\j@insplitin
\newwrite\j@insplitout
\newwrite\psbj@inaux
\immediate\openout\psbj@inaux=psbjoin.aux
\immediate\write\psbj@inaux{\string\joinfiles}%
\immediate\write\psbj@inaux{\jobname,}%
%
%
\def\toother#1{\ifcat\relax#1\else\expandafter%
  \toother@ux\meaning#1\endtoother@ux\fi}%
\def\toother@ux#1 #2#3\endtoother@ux{\def\tmp{#3}%
  \ifx\tmp\@mpty\def\tmp{#2}\let\next=\relax%
  \else\def\next{\toother@ux#2#3\endtoother@ux}\fi%
\next}%
%
%
\let\readfilenamehook=\relax
\def\re@d{\expandafter\re@daux}
\def\re@daux{\futurelet\nextchar\stopre@dtest}%
\def\re@dnext{\xdef\lastreadfilename{\lastreadfilename\nextchar}%
  \afterassignment\re@d\let\nextchar}%
\def\stopre@d{\egroup\readfilenamehook}%
\def\stopre@dtest{%
  \ifcat\nextchar\relax\let\nextread\stopre@d
  \else
    \ifcat\nextchar\space\def\nextread{%
      \afterassignment\stopre@d\chardef\nextchar=`}%
    \else\let\nextread=\re@dnext
      \toother\nextchar
      \edef\nextchar{\tmp}%
    \fi
  \fi\nextread}%
\def\readfilename{\bgroup%
  \let\\=\backslashother \let\%=\percentother \let\~=\tildeother
  \let\#=\sharpother \xdef\lastreadfilename{}%
  \re@d}%
%
%
\xdef\GlobalInputList{\jobname}%
\def\psnewinput{%
  \def\readfilenamehook{
    \if\matchexpin{\GlobalInputList}{, \lastreadfilename}%
    \else\xdef\GlobalInputList{\GlobalInputList, \lastreadfilename}%
      \immediate\write\psbj@inaux{\lastreadfilename,}%
    \fi%
    \let\readfilenamehook=\relax%
    \ps@ldinput\lastreadfilename\relax%
  }\readfilename%
}%
\expandafter\ifx\csname @@input\endcsname\relax    
  \immediate\let\ps@ldinput=\input\def\input{\psnewinput}%
\else
  \immediate\let\ps@ldinput=\@@input
  \def\@@input{\psnewinput}%
\fi%
\def\nowarnopenout{%
 \def\warnopenout##1##2{%
   \readfilename##2\relax
   \message{\lastreadfilename}%
   \immediate\openout##1=\lastreadfilename\relax}}%
\def\warnopenout#1#2{%
 \readfilename#2\relax
 \def\t@mp{TrashMe,psbjoin.aux,psbjoint.tex,}\uncatcode\t@mp
 \if\matchexpin{\t@mp}{\lastreadfilename,}%
 \else
   \immediate\openin\pst@mpin=\lastreadfilename\relax
   \ifeof\pst@mpin
     \else
     \edef\tmp{{If the content of this file is precious to you, this
is your last chance to abort (ie press x or e) and rename it before
retexing (\jobname). If you're sure there's no file
(\lastreadfilename) in the directory of (\jobname), then go on: I'm
simply worried because you have another (\lastreadfilename) in some
directory I'm looking in for inputs...}}%
     \errhelp=\tmp
     \errmessage{I may be about to replace your file named \lastreadfilename}%
   \fi
   \immediate\closein\pst@mpin
 \fi
 \message{\lastreadfilename}%
 \immediate\openout#1=\lastreadfilename\relax}%
{\catcode`\%=12\catcode`\*=14
\gdef\splitfile#1{*
 \readfilename#1\relax
 \immediate\openin\j@insplitin=\lastreadfilename\relax
 \ifeof\j@insplitin
   \message{! I couldn't find and split \lastreadfilename!}*
 \else
   \immediate\openout\j@insplitout=TrashMe
   \message{< Splitting \lastreadfilename\space into}*
   \loop
     \ifeof\j@insplitin
       \immediate\closein\j@insplitin\n@teoffalse
     \else
       \n@teoftrue
       \executeinspecs{\global\read\j@insplitin to\spl@tinline\expandafter
         \ch@ckbeginnewfile\spl@tinline
       \ifc@ntrolline
       \else
         \toks0=\expandafter{\spl@tinline}*
         \immediate\write\j@insplitout{\the\toks0}*
       \fi
     \fi
   \ifn@teof\repeat
   \immediate\closeout\j@insplitout
 \fi\message{>}*
}*
\gdef\ch@ckbeginnewfile#1
 \def\t@mp{#1}*
 \ifx\@mpty\t@mp
   \def\t@mp{#3}*
   \ifx\@mpty\t@mp
     \global\c@ntrollinefalse
   \else
     \immediate\closeout\j@insplitout
     \warnopenout\j@insplitout{#2}*
     \global\c@ntrollinetrue
   \fi
 \else
   \global\c@ntrollinefalse
 \fi}*
\gdef\joinfiles#1\into#2{*
 \message{< Joining following files into}*
 \warnopenout\j@insplitout{#2}*
 \message{:}*
 {*
 \edef\w@##1{\immediate\write\j@insplitout{##1}}*
\w@{
\w@{
\w@{
\w@{
\w@{
\w@{
\w@{
\w@{
\w@{
\w@{
\w@{\string\input\space psbox.tex}*
\w@{\string\splitfile{\string\jobname}}*
\w@{\string\let\string\autojoin=\string\relax}*
}*
 \expandafter\tre@tfilelist#1, \endtre@t
 \immediate\closeout\j@insplitout
 \message{>}*
}*
\gdef\tre@tfilelist#1, #2\endtre@t{*
 \readfilename#1\relax
 \ifx\@mpty\lastreadfilename
 \else
   \immediate\openin\j@insplitin=\lastreadfilename\relax
   \ifeof\j@insplitin
     \errmessage{I couldn't find file \lastreadfilename}*
   \else
     \message{\lastreadfilename}*
     \immediate\write\j@insplitout{
     \executeinspecs{\global\read\j@insplitin to\oldj@ininline}*
     \loop
       \ifeof\j@insplitin\immediate\closein\j@insplitin\n@teoffalse
       \else\n@teoftrue
         \executeinspecs{\global\read\j@insplitin to\j@ininline}*
         \toks0=\expandafter{\oldj@ininline}*
         \let\oldj@ininline=\j@ininline
         \immediate\write\j@insplitout{\the\toks0}*
       \fi
     \ifn@teof
     \repeat
   \immediate\closein\j@insplitin
   \fi
   \tre@tfilelist#2, \endtre@t
 \fi}*
}%
\def\autojoin{%
 \immediate\write\psbj@inaux{\string\into{psbjoint.tex}}%
 \immediate\closeout\psbj@inaux
 \expandafter\joinfiles\GlobalInputList\into{psbjoint.tex}%
}%
%
%
%
\def\centinsert#1{\midinsert\line{\hss#1\hss}\endinsert}%
\def\psannotate#1#2{\vbox{%
  \def\ps@nnotation{#2\global\let\ps@nnotation=\relax}#1}}%
\def\pscaption#1#2{\vbox{%
   \setbox\drawingBox=#1
   \copy\drawingBox
   \vskip\baselineskip
   \vbox{\hsize=\wd\drawingBox\setbox0=\hbox{#2}%
     \ifdim\wd0>\hsize
       \noindent\unhbox0\tolerance=5000
    \else\centerline{\box0}%
    \fi
}}}%
%
\def\at(#1;#2)#3{\setbox0=\hbox{#3}\ht0=0pt\dp0=0pt
  \rlap{\kern#1\vbox to0pt{\kern-#2\box0\vss}}}%
%
\newdimen\gridht \newdimen\gridwd
\def\gridfill(#1;#2){%
  \setbox0=\hbox to 1\pscm
  {\vrule height1\pscm width.4pt\leaders\hrule\hfill}%
  \gridht=#1
  \divide\gridht by \ht0
  \multiply\gridht by \ht0
  \gridwd=#2
  \divide\gridwd by \wd0
  \multiply\gridwd by \wd0
  \advance \gridwd by \wd0
  \vbox to \gridht{\leaders\hbox to\gridwd{\leaders\box0\hfill}\vfill}}%
%
\def\fillinggrid{\at(0cm;0cm){\vbox{%
  \gridfill(\drawinght;\drawingwd)}}}%
%
%
\def\textleftof#1:{%
  \setbox1=#1
  \setbox0=\vbox\bgroup
    \advance\hsize by -\wd1 \advance\hsize by -2em}%
\def\textrightof#1:{%
  \setbox0=#1
  \setbox1=\vbox\bgroup
    \advance\hsize by -\wd0 \advance\hsize by -2em}%
\def\endtext{%
  \egroup
  \hbox to \hsize{\valign{\vfil##\vfil\cr%
\box0\cr%
\noalign{\hss}\box1\cr}}}%
%
\def\frameit#1#2#3{\hbox{\vrule width#1\vbox{%
  \hrule height#1\vskip#2\hbox{\hskip#2\vbox{#3}\hskip#2}%
        \vskip#2\hrule height#1}\vrule width#1}}%
\def\boxit#1{\frameit{0.4pt}{0pt}{#1}}%
\catcode`\@=12 
%
\psfordvips   

\begin{flushright}
CERN-TH.7263/94.\\
LPTHE Orsay-94/49\\
HUTP-94/A015\\
HD-THEP-94-20\\
FTUAM-94/14\\
NSF-ITP-94-65\\
hep-ph/9406289
\end{flushright}
\newcommand{\be}{\begin{equation}}
\newcommand{\ee}{\end{equation}}
\newcommand{\bea}{\begin{eqnarray}}
\newcommand{\eea}{\end{eqnarray}}
\newcommand{\len}{\lefteqn}
\newcommand{\nn}{\nonumber}
\newcommand{\muh}{\hat\mu}
\newcommand{\dlr}{\stackrel{\leftrightarrow}{D} _\mu}
\newcommand{\vnew}{$V^{\rm{NEW}}$}
\newcommand{\vecp}{$\vec p$}
\newcommand{\dof}{{\rm d.o.f.}}
\newcommand{\prd}{Phys.Rev. \underline}
\newcommand{\pl}{Phys.Lett. \underline}
\newcommand{\prl}{Phys.Rev.Lett. \underline}
\newcommand{\np}{Nucl.Phys. \underline}
\newcommand{\vvp}{v_B\cdot v_D}
\newcommand{\dl}{\stackrel{\leftarrow}{D}}
\newcommand{\dr}{\stackrel{\rightarrow}{D}}
\newcommand{\mev}{{\rm MeV}}
\newcommand{\gev}{{\rm GeV}}
\newcommand{\calp}{{\cal P}}
\newcommand{\pinc}{\vec p \hskip 0.3em ^{inc}}
\newcommand{\pout}{\vec p \hskip 0.3em ^{out}}
\newcommand{\ptr}{\vec p \hskip 0.3em ^{tr}}
\newcommand{\pbr}{\vec p \hskip 0.3em ^{br}}
\newcommand{\no}{\noindent}
\newcommand{\ra}{\rightarrow}
\newcommand{\intsumpm} {\sum_{n^\pm}\hskip -14 pt\int\hskip 7 pt}
\newcommand{\intsump} {\sum_{n^+}\hskip -14 pt\int\hskip 7 pt}
\newcommand{\intsumm} {\sum_{n^-}\hskip -14 pt\int\hskip 7 pt}
\let\w=\omega
\let\G=\Gamma
\let\g\gamma
\let\X=\chi
\def\jor#1WAS#2{{%
\vrule width 1ex\raise-2pt\rlap{\vrule height0.4pt width2cm depth0pt}%
\bf#1}}
%
\def\dsl#1{\mathchoice
 {\dslaux\displaystyle{#1}} {\dslaux\textstyle{#1}} {\dslaux\scriptstyle{#1}}
 {\dslaux\scriptscriptstyle{#1}} }
\def\dslaux#1#2{\setbox0=\hbox{$#1{#2}$}
 \rlap{\hbox to \wd0{\hss$#1/$\hss}}\box0}
\let\eps\epsilon
\newcommand{\la}{\sin \theta_C\,\alpha_W}
\let\slash=\dsl
\def\Im{{\rm Im}}
\def\Re{{\rm Re}}

\pagestyle{empty}

\centerline{\LARGE{\bf{Standard Model CP-violation and  }}}
\vskip 1 cm
\centerline{\LARGE{\bf{Baryon asymmetry}}}
\vskip 1 cm
\centerline{\LARGE{\bf{Part II: Finite Temperature}}}

\vskip 1.5cm
\centerline{\bf{M.B. Gavela$^a$, P. Hernandez$^b$, J. Orloff$^c$,
O.P\`ene$^d$, C. Quimbay$^e$.}}
\centerline{$^a$ CERN, TH Division, CH-1211, Geneva 23, Switzerland}
\centerline{$^b$ Lyman lab., Harvard University, Cambridge, MA
02138\footnote{Junior Fellow, Harvard Society of Fellows}}
\centerline{$^c$ Institut f\"ur Theoretische Physik, Univ.
Heidelberg}
\centerline{$^d$ LPTHE, F 91405 Orsay, France,\footnote {Laboratoire
associ\'e au Centre National de la Recherche Scientifique.}}
\centerline{$^e$ CERN, TH Division, CH-1211, Geneva 23, Switzerland\footnote{On
leave of absence from Dpto. de F\'{\i}sica Te\'orica, Univ. Aut\'onoma de
Madrid, Cantoblanco, 28049 Madrid, and Centro Internacional de F\'{\i}sica,
Bogot\'a.}.}

\date{}
\begin{abstract}

 We consider the scattering of quasi-particles off the boundary created during
a first order electroweak phase transition. Spatial coherence is lost due to
the quasi-quark damping rate, and we show that reflection on the boundary is
suppressed, even at tree-level. Simply on CP considerations, we argue against
electroweak baryogenesis in the Standard Model via the charge transport 
mechanism. A CP asymmetry
is produced in the reflection properties of quarks and antiquarks hitting the
phase boundary.   An effect is present at
order $\alpha_W^2$ in rate and a regular GIM behaviour is found, which can be
expressed in terms of two unitarity triangles. A crucial role is played by the
damping rate of quasi-particles in a hot plasma, which is a relevant scale
together with $M_W$ and the temperature. The effect is many orders of
magnitude below what observation requires.

   \end{abstract}
\newpage
\pagestyle{plain}
\seCtion{Introduction}
A very concise summary of our main ideas and results on Standard Model (SM)
baryogenesis
in the presence of a first order phase transition, was recently presented in
ref.\cite{letter}.
 An academic\footnote{ Academic because the physical first order phase
transition is a thermal effect. In ref. \cite{nous0} we considered an
hypothetical $T=0$ world with two phases of spontaneous symmetry breaking
separated by a thin wall, to sharpen our tools and disentangle the effects
specific to the presence of a wall  from the  pure thermal ones.}  scenario at
zero temperature  is developed with all the particulars in ref.\cite{nous0}.
The aim of the present work is to consider in detail the  finite temperature
($T$) scenario, where decoherence effects become essential.

  Gamma-ray and cosmic data do not show any evidence for primary antiparticles
on scales up to the level of clusters of galaxies. Nucleosynthesis constraints
require a baryon number to entropy ratio in the observed part of the universe
$n_B / s \sim (4-6) 10^{-11}$ at $T > 1$ GeV \cite{exp}. In the absence of a
sensible mechanism for matter-antimatter separation on such large scales, a
plausible alternative is to require that the microscopic laws of physics are
responsible for such a baryonic excess\cite{sak}. This scenario is called
baryogenesis.

The SM electroweak phase transition occurs rather late in the cosmological
evolution, at a time when the expansion of the universe is slow compared to
weak interaction time scales and $T\sim100$ GeV.   Sakharov's
condition\cite{sak} of departure from thermal equilibrium could be fulfilled
through a sudden, first order, phase transition\cite{trans0}\cite{trans}. In
this case, bubbles of true ground state, characterised by a non-zero vacuum
expectation value of the Higgs field $v$, grow and fill the preexisting $v=0$
universe. The evident non-equilibrium element is given by the propagation of
the bubble surfaces, and physics in their vicinity should be examined in view
of baryogenesis\cite{trans2}.

It has been known as well for many years that weak interactions  violate baryon
number\cite{spha}. At low temperatures, the rate is negligible, while at high
temperatures\cite{dimo} the rate is enhanced,
\be
\Gamma^{u}_{sph} \sim \kappa (\alpha_W T)^4 ,\qquad\qquad\qquad v=0
\label{ratein}\ee
\be \Gamma^{b}_{sph} \sim \exp\left [-\frac{M_W(T)}{\alpha_W T}\right
],\qquad\qquad v\ne 0
\label{rateout}\ee
where $v$ denotes the vacuum expectation value of the Higgs field, and $\kappa$
is a parameter of $O(1)$ or smaller.

The SM complies as well with the C and CP violation requirements. In particular
 the presence of complex Yukawa couplings induces a  phase in the
Cabibbo-Kobayashi-Maskawa (CKM)\cite{KM} matrix for three generations,
responsible for CP violation.

\begin{figure}[hbt]
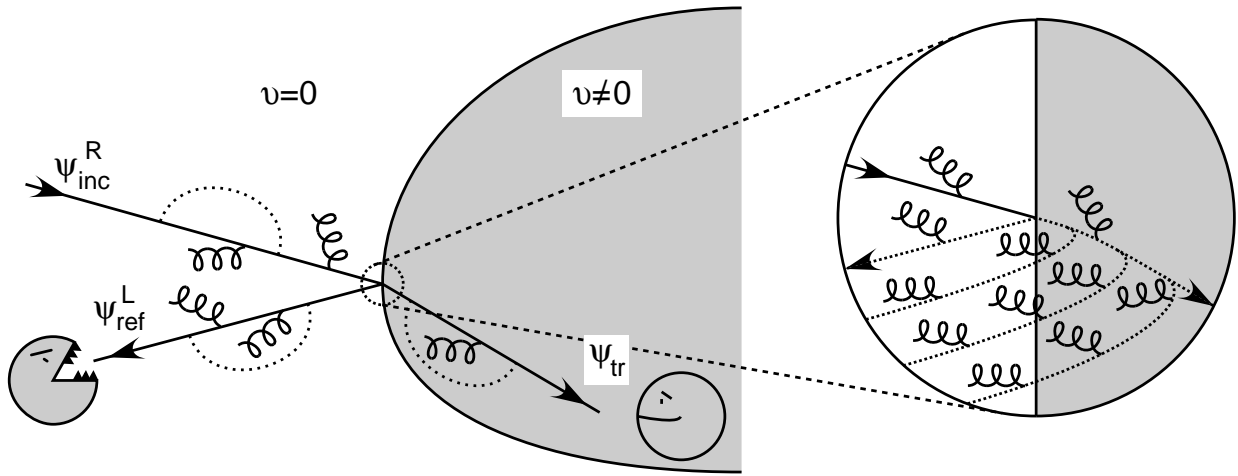
   
    \begin{center} \mbox{\psboxto(\hsize;0pt){pacmanii.ai}} \end{center}
      \caption{\it Artistic view of the charge transport mechanism, as
      described in the text. The hungry ``pacman'' represents rapid sphalerons
      processes. The wiggly lines stand for collisions with thermal
      gluons. Only electroweak loops are depicted,
      represented by dotted lines.}  \protect\label{pacman}
\end{figure}

 Sakharov's conditions imply that both baryon number-violating and C and
CP-violating processes have to undergo an out-of-equilibrium period.  The
currently proposed scenario for SM baryogenesis\cite{shapo} is a charge
transport mechanism \cite{ckn}. It is graphically summarised in
Fig. \ref{pacman}, where we have zoomed into the vicinity of one of the
bubbles. In the wall rest frame a net baryonic flux is hitting the surface
from the unbroken phase. Consider for definiteness the trajectory of an
initial right-handed quark. Upon eventual reflection, a
left-handed quark will travel backwards to the unbroken
($v=0$) phase.  Electroweak interactions have been {\it a priori} active
throughout the trajectory; their CP-violating effects induce a different
reflection probability for quarks and antiquarks and, thus, a CP asymmetry in
the reflected flux. Then, unsuppressed sphaleron processes,
eq. (\ref{rateout}), restoring equilibrium in the unbroken phase, can
``swallow'' the outgoing quarks, transforming the CP asymmetry
into a baryonic one. The sweeping of the expanding bubble will automatically
transfer the latter to the broken phase ($v\ne0$), where we live. An important
survival requirement for the produced baryon asymmetry was given in
refs. \cite{higgs1} and \cite{linde2}, by requiring that sphaleron processes
inside the bubble are weak enough so as not to wash it out. In perturbation
theory, this results in an upper bound of the Higgs mass $\sim 45$ GeV, in
conflict with experiment.  This is a big problem for the scenario, although
the perturbative treatment may be inadequate, and the question is the subject
of much work at present.

We will not enter the discussion on whether a first order phase transition did
take place. It will be assumed that it did, and that an optimal sphaleron rate
is present as well. Our aim is to argue, on a quantitative estimation of the
electroweak C and CP effects exclusively, that the current SM scenario is
unable to explain the above mentioned baryon number to entropy ratio\footnote{A
discussion of intuitive expectations is given in the introduction of
ref.\cite{nous0}}. Notice that once a CP-asymmetry on the reflected baryonic
current, $\Delta_{CP}$, is obtained, the induced baryon asymmetry is at most
$n_B/s\,\sim\,10^{-2} \Delta_{CP}$, in a very optimistic estimation of the
non-CP ingredients\cite{linde2}\cite{shapo}. $\Delta_{CP}\ge 10^{-8}$ is thus
required.

 The symmetries of the problem are analyzed in detail for a generic bubble in
ref.\cite{nous0}.
 Major unknowns of the scenario are the wall velocity, $\vec v_{wall}$, and the
wall thickness, $l$. Typical values for $|\vec v_{wall}|$  are
non-relativistic, $\sim 0.1-0.4$\cite{linde2}\cite{vwall}.
 Our analytical results correspond to the thin wall scenario. The latter
provides an adequate physical description for typical momentum of the incoming
particles  $\vert \vec p \vert \ll 1/l$. For higher momenta the cutoff effects
would show up, and indeed ``realistic'' walls have $l\sim 10/T\,-\,20/T$, but
it is reasonable to believe that the thin wall approximation produces an upper
bound for the CP asymmetry. We work in a simplified scenario with just one
spatial direction, perpendicular to the wall surface: phase space effects in
the $3+1$ dimension case would further suppress the effect. It is assumed as
well that both phases are in thermal equilibrium : this is only broken by the
wall motion.

At $T\ne 0$ the correct incoming asymptotic states are quasiparticles instead
of particles, built up of a resummation of the thermal self-energies of the
particles. Nevertheless, the three building blocks of the problem are
essentially the same as in the $T=0$ case\cite{nous0}, to wit:
the CP-violating couplings of the CKM matrix, the presence of CP-even phases
associated to complex reflection coefficients
 for certain values of the energy of the incoming quasi-particles, and the
well-known fact that, at $T \ne 0$, the fermionic on-shell self-energy cannot
be
completely renormalized away and induces physical transitions already at the
one-loop level. Even starting from massless
particles, the eigenstates of the effective thermal Dirac equation have an
``effective mass'', or plasma frequency, due to the QCD thermal self-energies.
It gives
 the overall energy
scale of the problem $\sim g_sT\sim 50$ GeV, for $T\sim 100$ GeV. This
results in a shift in the position of the reflection coefficient threshold for
a given quasi-particle. Nevertheless, the tree-level reflection region for a
given flavour is still of the order of the corresponding current (non-thermal)
mass  $m$. In other words, the time required for complete reflection of a given
quasi-particle is of the order of $1/m$.

On top of the above, the authors of ref.\cite{shapo} have pointed out that
electroweak thermal loops are present as well, and because the latter are
flavour dependent, their resummation can lift the QCD degeneracy of the
spectrum of quasi-quarks even far from the wall in the unbroken
 phase\footnote{Unlike the $T=0$ case,
 Lorentz invariance is lost
 even far from (without)
the wall. At $T=0$, far from the wall
 in the unbroken phase,  all
particles are massless, at any order in the self-energy corrections.
  Any base is a good one, as rotations
are physically irrelevant.} . The
result would be further shiftings and reshufflings in the reflection
coefficient thresholds\footnote{We will see that, finally, it just amounts to
technical complications.} and, more important, a sizable CP asymmetry due to
interference between different flavours and chiralities. In our opinion, the
important point to retain is that CP violation is a quantum phenomenon, and
can only be observed when quantum coherence is preserved over time scales
larger than or equal to the electroweak time scales needed for CP violation.

We argue\cite{letter} that the quantum phase of the quasi-quarks is in fact
lost much before the time scales mentioned in the above paragraphs. A further
fundamental difference with the $T=0$ case is the damping rate, $\gamma$, of
quasi-particles in a plasma. Small momenta are relevant for the problem under
study, and it is known that at zero momentum, the QCD contribution to the
damping rate is $\gamma\sim 0.15\,g_s^2\,T$, i.e. $\sim 19$ GeV.  Due to
incoherent thermal scattering with the medium, the quasi-particle has a finite
life-time $\sim 1/2\gamma$, turning eventually into a new state, out of phase
with the initial one. Although the life-time is larger than the overall QCD
time scale mentioned above, $\sim 1/(50$GeV), it is small in comparison with
both the electroweak  thermal self-energy time scales for any quasi-quark, and
with the tree-level reflection times $\sim 1/m$ for all quasi-quarks but the
top.

 The problem can be rephrased in terms of spatial decoherence. Reflection is a
spatial property. The quasi-quarks have a group velocity of $\sim 1/3$ and thus
a mean free path, or coherence length, of $\sim 1/6\gamma$. Over larger scales,
such as those needed for instance for total reflection, the quantum coherence
of light quasi-quarks is damped and any pure quantum effect, such as CP
violation, is suppressed.

 We thus show that tree-level reflection is suppressed for any light flavour by
a factor $\sim \,m/2\gamma$. The presently discussed CP-violation observable
results from the convolution of this reflection effect with electroweak loops
in which the three generations must interfere coherently in order to produce a
CP-violation observable. It follows that further factors of this type appear
in the final result, which is many orders of magnitude below what observation
requires and has an ``\`a la Jarlskog''\cite{Jarlskog} type of GIM
cancellations. We show as well that the effect is present at order
$\alpha_W^2$ in a perturbative expansion, as in the $T=0$ case\cite{nous0}.

The scope of the present paper goes beyond the particular issue of baryon
number generation in the SM. Spatial loss of quantum coherence may be relevant
in other microscopic processes at finite temperature. A more rigorous
treatment than the heuristic one developed here would be welcome, though. In
our opinion, this general topic deserves further attention.

In Sect. \ref{sec-ctm} we discuss on general grounds the relation between
baryonic and
 CP asymmetries. Sect. \ref{sec-spec} describes the quasi-particle spectrum far
from the wall both in the unbroken and the broken phase, as well as at the
interface. It contains the dispersion relations, exemplified by the
corresponding effective Lagrangians/Hamiltonians which take into account the
QCD damping rate. Sect. \ref{sec-wall} considers  the tree-level scattering of
a quasi-quark hitting the bubble wall from the unbroken phase in a one-flavour
world. It parametrises how the loss of quantum coherence results in a damping
of the reflection properties of the plasma. A density matrix formalism is
developed as well. Sect. \ref{sec-CP} considers the same scattering problem in
the realistic case of three generations with mixing. It includes the
computation of the CP asymmetry as well as comparison with literature and
comments on wall thickness effects. The conclusions are summarised in Sect.
\ref{sec-conclu}.

\seCtion{Baryon Asymmetry Through Charge Transport}
\label{sec-ctm}
In the introduction, we have discussed qualitatively how it is possible
to generate a baryon asymmetry in the context of
 a first order phase
transition by fermion transport. In this section, we state more
precisely the relation between the baryon asymmetry, $n_b/s$, and the  CP-odd
asymmetry that we aim to compute.

In the fermion transport mechanism, we can distinguish two steps.
Firstly,
a spatial \underline{chiral} baryon number separation is produced due to the
CP-violating scattering of quarks on the
wall.

Denote by $r^{u,b}_\chi$, $t^{u,b}_\chi$ (${\bar{r}}^{u,b}_\chi$,
${\bar{t}}^{u,b}_\chi$), the reflection and transmission amplitudes from the
unbroken ($u$) or broken ($b$) phases, for an incoming quark (antiquark) of
chirality $\chi$ \footnote{For anti-quarks, the CP conjugate of a quark of
chirality $\chi$ has chirality $-\chi$. More precisely, the doublet quark
(antiquark) has chirality $-1$ ($+1$), and conversely for singlets. To
simplify notations, $\chi$ will indifferently stand for ($L,R$) or the
associated $\gamma_5$ eigenvalue (-1,+1).}.  In our convention, a positive
velocity is defined as flowing from the unbroken to the broken phase.  The
spin $\sigma_z$ is a conserved quantum number. For positive (negative) group
velocity, its eigenvalue in the unbroken phase is given by chirality
$+(-)\chi$, as will be shown in section (\ref{sec-one}). In the broken phase,
although chirality is no longer a good quantum number, we keep the same
notation for the $\sigma_z$ eigenvalues.  The conserved current density in the
$z$-direction is
\be j_z \propto \bar u(p')  \gamma_z u(p) = V_z u^\dagger(p') u(p), \, for\,\,p=p'\label{velo}\ee
where $\vec{V}$ is the group velocity. For convenience we will extend eq. (\ref{velo}) to $p \ne p'$, in which case it may be taken as a definition of a ``non diagonal'' $V_z$. 
For an incoming spinor $u^u_\X$, normalised as ${u^u_\X}^\dagger u^u_\X = 1$,
we have
\bea
\frac{j_{ref}}{j_{inc}} = |r^u_\X|^2 \,|V^u_{-\X}/V^u_{\X}| \,=
|r^u_\X|^2,\qquad
\frac{j_{tr}}{j_{inc}} = |t^u_\X|^2 \, |V^b_\X/V^u_{\X}|,
\eea
where we used $|V^u_{-\X}/V^u_{\X}| = 1$, consistent with our future
approximations.  The one particle thermal density matrix in the wall rest
frame, $\rho^{u,b}_\chi$, is a positive definite hermitian matrix
\footnote{The density matrix for antiquarks of chirality $-\chi$ is the same,
if we assume there are no primeval asymmetries.}. It is not translationally
invariant (i.e. it is non-diagonal in momentum), due to the presence of the
wall and depends on the plasma velocity, $-v_{wall}$.  Also, it is different
for doublet and singlet quarks, as electroweak interactions generate an
$O(\alpha_w)$ difference in their thermal self-energies.

The density matrix of incoming particles for one given chirality $\X$ is
normalised as usual,
\be  \int dp \;\mbox{Tr}[\rho^u_\X(p,p)]=1, \ee
where Tr stands for the trace in flavour space. This corresponds to having one
particle in the total volume $L$, i.e. to a particle density far from the wall
of $1/L$.

The chiral baryon number current density in the unbroken phase, for a unit
incoming density, is then given by the sum of the corresponding reflected and
transmitted chiral current densities\footnote{In deriving \ref{lastd} we have
used the assumption (trivially verified for non-interacting particles
prepared in equilibrium in a half space) that
$\rho^{u}_{-\chi}(p,p')\, r^u_{-\chi}(p'){{r^u}_{-\chi}(p)}^\dagger + $
$\rho^{b}_{\chi}(p,p')\, t^b_{\chi}(p'){{t^b}_{\chi}(p)}^\dagger$ $
|V_\X^u/\sqrt{|V_\X^b(p)V_\X^b(p')|}$ is the matrix density for the particles
of chirality $\X$ going away from the wall in the unbroken phase, and
analogously for antiquarks. The full density matrix further has
chirality non diagonal pieces, but these do not contribute to the computed
current.},
\bea\len{
\Delta^u_\chi \equiv \frac L {2\pi}\int\int dp dp' \mbox{Tr}[\,
\rho^{u}_{-\chi}(p,p')\, \{ r^u_{-\chi}(p'){{r^u}_{-\chi}(p)}^\dagger
-{{\bar{r}}^u_{\chi}}(p'){{\bar{r}}^u_{\chi}(p)}^\dagger\} ]
\nonumber}\\
&+ & |V^u_\X|/\sqrt{|V^b_{\X}(p) V^b_{\X}(p')|}
\,\mbox{Tr}[\,\rho^{b}_{\chi}(p,p')\, \{
t^b_{\chi}(p'){{t^b}_{\chi}(p)}^\dagger -
{{\bar{t}}^b_{-\chi}}(p'){{{\bar{t}}}^b_{-\chi}(p)}
^\dagger\} ].\label{lastd}
\eea

This expression can be simplified by using the relations imposed by unitarity
and CPT symmetry, or more exactly CP'T. The latter is defined \cite{nous0} as
the combination of CPT and a $\pi$ rotation around the $y$ axis, which leaves
the wall invariant.

\begin{itemize}
\item CP'T
\bea
r_{\chi}^u = \left({\bar{r}}^u_{\chi}\right)^{t},\;\;\; r^b_{\chi} =
\left({\bar{r}}^b_{\chi}\right)^{t},\;\;\;
t^u_{\chi} = \left(\bar{t}^b_{-\chi}\right)^{t}
\eea
where superscript $t$ stands for transposition in flavour space.
\item Unitarity, i.e. current conservation,
\bea
 r^u_{\chi}{r^u_{\chi}}^\dagger + |V^b_\X/V^u_{\X}|\,
t^u_{\chi}{t^u_{\chi}}^\dagger = 1 ,\qquad r^b_{\chi}{r^b_{\chi}}^\dagger +
|V^u_\X/V^b_{\X}|\, t^b_{\chi}{t^b_{\chi}}^\dagger = 1.
\eea
\end{itemize}

Then,
\bea
\Delta^u_{\chi} = \frac L {2\pi} \int \int dp\,dp'\,\mbox{Tr}[(\rho^{u}_{-\chi}
- \rho^{b}_{\chi})
\{  r^u_{-\chi}{(r^u_{-\chi})}^\dagger -
{{\bar{r}}^u_{\chi}}{({\bar{r}}^u_{\chi})}^\dagger\} ]
\label{asy}
\eea
where the possible flavour non-diagonal elements in $(\rho^{b}_{\chi})^{t}$
have been neglected, and the $p, p'$ dependence omitted for simplicity.

The total outgoing baryon number current density  in the unbroken phase, per
unit incoming quark/antiquark flux, is:
\bea
B^u = \Delta^u_{\chi} +  \Delta^u_{-\chi} =  \frac L {2\pi} \int \int dp\,dp'\,
\mbox{Tr}[ (\rho^{u}_{-\chi} - \rho^{u}_{\chi} + \rho^{b}_{-\chi} -
\rho^{b}_{\chi} )
\{  r^u_{-\chi}{(r^u_{-\chi})}^\dagger - {{\bar{r}}^u_{\chi}}
{({\bar{r}}^u_{\chi})}^\dagger\}] \eea

For $\rho^{u,b}_{-\chi} = \rho^{u,b}_{\chi}$, clearly $B^u = 0$. This shows
 that, up to order $O(\alpha_w)$, only {\it chiral} baryon number gets
 separated by the wall.  We will neglect this subleading effect and
 assume $\rho^{u,b}_{-\chi} =
\rho^{u,b}_{\chi} =\rho^{u,b}$ from now on.

Furthermore, neglecting the current (non-thermal) mass  effects in the thermal
distributions \footnote{This is
a good approximation for quark masses smaller than $g_s T$.}, $\rho^u$ and
$\rho^b$ only differ because of the opposite boost. If  $\rho^{u,b}$ were the
boosted Fermi thermal distribution, and for small
wall velocities,
\be
\rho^u - \rho^b \simeq v_{wall} f \rho.
\label{rhodif}
\ee
where $\rho$ corresponds to the Fermi thermal distribution in the plasma rest
frame, $\rho \propto n_F$, and $f = \rho^{-1}\partial \rho /\partial v^{wall} <
\,1$ \footnote{ For the  Fermi distribution, $f=(1-n_F)\frac {p^u-p^b}T$, where
 $p^u\simeq p,p'$ is the momentum in the unbroken phase, and  $p^b$ in the
broken one.}. For the problem under study, $\rho$ differs
from the Fermi distribution, but  the relation (\ref{rhodif}), with $f<1$,
should still be a good order of magnitude estimate.

In this approximation, we
define the CP asymmetry as
\be
\Delta_{CP} = \frac L {2\pi}\mbox{sign}(\chi) \int \int
dp\,dp'\,\mbox{Tr}[\;\rho\;
\{  r^u_{-\chi}{(r^u_{-\chi})}^\dagger -
{{\bar{r}}^u_{\chi}}{({\bar{r}}^u_{\chi})}^\dagger\} ].
\label{dcp}
\ee

In a second step, the sphaleron transitions are taken into account. We do not
 consider strong sphalerons \cite{stspha}, which would further reduce the effect
\cite{giudice}, as they require to take into account the $O(\alpha_W)$ difference in
 the thermal distributions for doublets and singlets mentioned above.     
  Weak sphalerons only
affect left chirality particles (or right antiparticles).  As explained in
the introduction, these sphaleron processes have a very different rate in
each phase, eqs. (\ref{ratein}) and (\ref{rateout}). In
an ideal picture, all the left baryon number in the unbroken phase is
completely diluted, while the one in broken phase remains.  Eventually, when
the phase transition is completed, the remaining total baryon number, divided
by the original number $n$ of quarks with one given chirality and spin, is
\bea
n^{optimal}_B/n = \frac 1 3 \Delta^u_R = -\frac 1 3\Delta^u_L \simeq \frac 1 3
v^{wall} \Delta_{CP}.
\label{nopt}
\eea
Only the contributions from the unbroken phase appear, as sphalerons
transitions are absent from the broken phase. $n^{optimal}_B/s$ is smaller, as
the total entropy $s$ is  at least one order of magnitude larger than
$n$\footnote{There are 4 spin-chiralities per quark and per antiquark, three
generations, and  additional quanta in the plasma (leptons and gauge bosons).
Besides, the quasiparticles correspond to a limited part of phase space, $E\sim
g_s T$.}.

Unfortunately, the sphaleron processes are not so efficient and computing
the actual number of sphaleron transitions is a difficult diffusion problem
. The authors of \cite{shapo} give an already optimistic estimate for $n_b/s$
which is $\sim 10^{-2} \Delta_{CP}$.
Without relying on this calculation, it is necessarily true that
\bea
\frac{n_b}{s} < \Delta_{CP}.
\eea
$\Delta_{CP}$ gives a highly conservative upper bound for $n_b/s$.
This will be enough to rule out SM baryogenesis within
this mechanism, and we will devote the rest of the paper to the computation of
this asymmetry.

\seCtion{The spectrum of quasi-particles.}
\label{sec-spec}

  In this section the spectrum of quasiparticles is derived from the  thermal
loop
contribution to the quark self-energy. The solutions in a world with just one
phase, either spontaneously broken ($v\ne 0$) or unbroken ($v=0$), are computed
first. Secondly, the case when the two phases coexist separated by a boundary
(thin wall) is discussed, as well as the matching of the two spectra.

\subsection{The quark self-energy.}
\label{sec-loop}

This is a rather technical subsection, whose results are needed later. We
present a novel calculation of the real part of the self-energy, including the
electroweak contributions, in the broken phase.
 The computations are performed in
the real time formalism. We compare to previous results for the unbroken phase.
 The imaginary part computed in QCD at zero momentum\cite{bp} is considered.

\subsubsection{Real part in the broken phase.}
\label{sec-real}

We  use the notations of \cite{petitg},
generalised to several flavours:

\be \Re(\Sigma(k))=-a \slash k- b \slash u - c m\label{petit} \ee
where $a$, $b$, $c$ and $m$ are Lorentz invariant matrices in flavour space,
$u$
is the four-velocity of the plasma and $k=(\omega, \vec k)$ is the external
momentum. $m$ denotes the mass matrix for the external flavours. In the plasma
rest frame and the mass basis,
\be
\Re(\Sigma(\omega,\vec k))\gamma_0 = -h(\omega,\vec k) - a(\omega,\vec k) \vec
\alpha \cdot
\vec k - c(\omega,\vec k) m\gamma_0 \label{resigma}\ee
where
\be h(\omega,\vec k)=a(\omega,\vec k)\,\omega+b(\omega,\vec k)
.\label{hleftb}\ee

The one loop calculation gives the following matrix elements:
\be a(\omega,k)_{fi}=f\,A(m_i,0)\delta_{fi}+
\frac{g^2}{2}[\sum_{l}f_{W,l}\,A(M_l,M_W)+
(f_Z\,A(m_i,M_Z)+f_H\,A(m_i,M_H))\delta_{fi}]\label{aa}\ee
where the index $l$ runs over the internal flavours for $W$ exchange, and $M$
denotes the corresponding mass. In
this expression,
\bea &f=[\frac {4}{3}g_s^2+Q_i^2g^2s_W^2](L+R)\quad,\quad
f_{W,l}=[(1+\frac{\lambda_l^2}{2})L+
\frac{\lambda_i\lambda_f}{2}R]K_{li}K^*_{lf}\nn\\
&f_Z=\frac{1}{2}(\frac{4}{c_W^2}(T_i^3-Q_i s^2_W)^2+
\frac{\lambda_i^2}{2})L+(\frac{4}{c_W^2}(-Q_i s^2_W)^2+
\frac{\lambda_i^2}{2})R\quad,\quad
f_H=\frac{\lambda_i^2}{4}(L+R),\label{fWZH}\eea
where $L, R$ are chiral projectors. $K$ represents the
Cabibbo-Kobayashi-Maskawa matrix and
$\lambda_i = m_i/M_W$ are related to the usual Yukawa couplings by the
following relation,
\be
f_i = \frac{g \lambda_i}{\sqrt{2}}.
\ee
The integral is given by:
\bea
 &A(M_F,M_B)= \frac{1}{k^2} \int^\infty _0 \frac{dp}{8\pi^2}
([-\frac{(\omega^2+k^2+\Delta)}{2k}
\frac{p}{E_B}L_I^{+}(p)-
\frac{\omega p}{k}L_I^{-}(p)+
\frac{4p^2}{E_B}]n_B(E_B)+\nn\\&\qquad\qquad\qquad
[\frac{\omega^2-k^2-\Delta}{2k}
\frac{p}{E_F}L_{II}^{+}(p)-
\frac{\omega p}{k}L_{II}^{-}(p)+
\frac{4p^2}{E_F}]n_F(E_F)),
\label{eqA}\eea
where

\be
L^{\pm}_I(p)= [\log\frac{2kp+2E_B
\omega+\omega^2-k^2+\Delta}{-2kp+2E_B \omega+\omega^2-k^2+\Delta}]
\pm [E_B\rightarrow -E_B]\label{LI}\ee
and

\be L^{\pm}_{II}(p)= -[\log\frac{2kp-2E_F
\omega+\omega^2-k^2-\Delta}{-2kp-2E_F \omega+\omega^2-k^2-\Delta}]
\mp [E_F\rightarrow -E_F]\label{LII},\ee
with
\be E_{F,B}=\sqrt{p^2+M_{F,B}^2},\,\,  \Delta=M_B^2-M_F^2,\,\,
n_{F,B}(E)=(\exp{E/T}\pm 1)^{-1}.\label{BF}\ee

The function $h(\omega,k)$ is given by

\be h(\omega,k)_{fi}=-f\,H(m_i,0)\delta_{fi}-
\frac{g^2}{2}[\sum_{l}f_{W,l}\,H(M_l,M_W)+
(f_Z\,H(m_i,M_Z)+f_H\,H(m_i,M_H))\delta_{fi}],\label{hh}\ee
where

\be
 H(M_F,M_B)=\frac{1}{k}
 \int^\infty_0
\frac{dp}{8\pi^2}([p L_I^{-}(p)+\frac{\omega p}{
E_B}L_I^{+}(p)]n_B(E_B)+p L_{II}^{-}(p)n_F(E_F)).
\label{eqH}\ee

The chirality breaking term $c(\omega,k)$ has the following expression
$$
c(\omega,k)_{fi}=2 f\,C(m_i,0)\delta_{fi}+
\frac{g^2}{2}
\left[\sum_{l}\frac{M_l}{m_i}g_{W,l}\,C(M_l,M_W)+\right. $$
\be 
(g_Z\,C(m_i,M_Z)-f_H\,C(m_i,M_H))\delta_{fi}\bigg],\label{ac}\ee
with

\bea
g_{W,l}=&\frac{\lambda_M}{2}(\lambda_f L +\lambda_i
R)K_{li}K^*_{lf},\quad
g_Z=\frac{1}{2}[-8 Q_i \frac{s^2_W}{c^2_W} (T_3 - Q_i s_W^2) +
\frac{\lambda_i^2}{2}](L+R)\label{gWZ}\eea
and
\bea
&C(M_F,M_B)=\frac{1}{k}
\int^\infty_0
\frac{dp}{8\pi^2}(\frac{p}{ E_B}L_I^{+}(p) n_B(E_B)+\frac{p}{E_F}
L_{II}^{+}(p)n_F(E_F)).
\label{more}\eea

\subsubsection{Real part in the unbroken phase.}
\label{sec-realun}

In the unbroken world the expressions (\ref{aa})-(\ref{more})
apply with all masses equal to zero. For the sake of comparison with previous
literature, we give here the leading ($O(T^2)$) contribution when $\omega,
\vert \vec k \vert \ll T$.
In this limit, the one loop fermionic self energy in the unbroken phase has
been computed in QCD \cite{klimov}, \cite{weldon} and generalised to
electroweak interactions in \cite{shapo}. We find:
\be
h(\omega,\vec k)_L=- \frac{T^2}{\omega}F\left(\frac {\omega}{\vert
k\vert}\right)
\left\{\frac{2\pi\alpha_s }{3}+\frac{3\pi\alpha_W }{8}\left(1+\frac{\tan^2
\theta_W}{27}+\frac 1 3 (\lambda_{i}^2+K^*_{lf} \lambda_{l}^2
K_{li})\right)\right\},\label{hleft}\ee

\be h(\omega,\vec k)_R=- \frac{T^2}{\omega}F\left(\frac {\omega}{\vert
k\vert}\right)
\left\{\frac{2\pi\alpha_s }{3}+\frac{\pi\alpha_W }{2}\left({Q_{i}^2\tan^2
\theta_W}+\frac 1 2 \lambda_{i}^2\right)\right\}\label{hright},\ee
with
\be
a(\omega,\vec k)_L=-h(\omega,\vec k)_L\,\,\frac{\omega \left(1-F\left(\frac
{\omega}{\vert k\vert}\right)\right)}{\vert k\vert^2 F\left(\frac
{\omega}{\vert k\vert}\right)},
 \label{aleft}\ee
and
\be a(\omega,\vec k)_R=-h(\omega,\vec k)_R\,\,\frac{\omega \left(1-F\left(\frac
{\omega}{\vert k\vert}\right)\right)}{\vert k\vert^2 F\left(\frac
{\omega}{\vert k\vert}\right)}.
 \label{aright}\ee
$a_{L,R}$, $h_{L,R}$ correspond to the coefficients of the
projectors $L,R$ in $a$,$h$, and
\be F(x)=\frac{x}{2}\left(\log\left(\frac{x+1}{x-1}\right)\right).
\label{fdex}\ee
In  eqs. (\ref{hleft})-(\ref{aright}) it is understood that all terms which do
not contain the CKM matrix are flavour diagonal\footnote{The careful reader may
notice several differences in numerical factors between eqs
(\ref{hleft})-(\ref{aright}) and eqs. (6.6)-(6.8) in \cite{shapo}, once  the
change of basis is taken into account.}.

\subsubsection{Imaginary part of the Self-energy.}

 It is known that the one loop calculation of the imaginary part of the self
energy (proportional to the damping rate) is incomplete. A leading order QCD
computation of the quasi-quark damping rate\cite{bp} has been performed at zero
$\vec k$: $\gamma \sim 0.15 g_s^2 T$  i.e. $\sim 19$ GeV at $T=100$ GeV. We
will neglect its electroweak component, and assume that in the vicinity of
$\vec k=0$ the damping rate remains close to the above-mentioned value, i.e.
\be \gamma_0\Im(\Sigma(\omega, \vec k)) \simeq -2\gamma\label{imb}\ee

\subsection{The spectrum.}
\label{sec-one}
The spectrum of quasi-particles in a quark-gluon plasma with only QCD
interactions
has been extensively studied in the literature \cite{klimov}, \cite{weldon}.
The effects
we need for generating a baryon asymmetry are  electroweak, though, and it is
necessary to include the electroweak thermal loops. The latter will modify the
spectrum in a
complicated way, although the corrections are quantitatively small with respect
to the QCD contributions.
This will allow a convenient perturbative treatment, as it will be shown.

Let us summarise first the results in pure QCD, i.e. setting $\alpha_w =0$
in the self-energies (\ref{resigma}). The spectrum of quasi-particles is given
by the
isolated poles of the propagator (or the zeros of the self-energy).
At $\vec{k} = 0$, it is given by the solution of
\be
\omega + h_{QCD}(\omega,0) - 2 i \gamma = 0
\label{pole}
\ee

To gain some insight into the energy scale of the spectrum, we simplify
(\ref{pole}) neglecting all the energy scales other than T (i.e. $m_i << T$).
We work to
first order in $\gamma/(g_s T)$, an approximation we will stick to in the rest
of the paper. The solution then is
\be
w^0_{QCD}\simeq T \sqrt{\frac {2 \pi \alpha_s}3}-i\gamma
\equiv \omega^0_{QCD} -i\gamma\label{omega0}.\ee

Eq. (\ref{omega0}) shows that QCD has shifted dramatically the position of the
singularities of the quark propagator (by singularity we mean the pole
and the location of the thermal $\delta(p^2-m^2)'s$), and sets the
energy scale of the quasi-particles at $O(g_s T)\sim 50$ GeV.

The dispersion relations $w(\vec k)_{QCD}$ satisfied by these quasi-particles
are given by the vanishing eigenvalues of the
matrix
 \be
i\gamma_0 S^{-1} ={\omega}+h_{QCD}(\omega,\vec k)
 -\gamma_5 (\vec \sigma \cdot \vec k)(1+a_{QCD}(\omega,\vec k)) -2 i \gamma.
\label{qcd}\ee

We are interested only in the low momentum region, and upon linearising in
momentum we find,
\be
w(\vec k)_{QCD} = \omega^0_{QCD} +\chi (\vec \sigma \cdot \vec k) V_{QCD}- i
\Gamma_{QCD},\label{disqcd}\ee
where $\chi$ is the eigenvalue of $\gamma_5$ (the chirality), i.e. +1(-1) for
right(left)-handed fermions, and
\be
V_{QCD} \equiv \frac{1+a_{QCD}(\omega^0_{QCD},0)}{
 s_{QCD}(\omega^0_{QCD},0)},\qquad
\Gamma_{QCD}=\frac{2\gamma}{s_{QCD}(\omega^0_{QCD},0)},
\label{alr}\ee
with
\be
s_{QCD}(\omega_{QCD}^0,0)=1+\frac{\partial
h_{QCD}(\omega,0)}{\partial \omega}\vert_{\omega^0_{QCD}}.
\label{est}\ee

The functions $s_{QCD}$ are the inverse of the residues at the poles.
In a second quantisation treatment of the quasi-particles, they can be
absorbed through wave function renormalisation, in order to properly normalise
the kinetic terms. When deriving eq. (\ref{disqcd})  higher orders in
$\gamma/(g_s T)$ have been consistently neglected, leading to real arguments in
eqs. (\ref{alr}) and (\ref{est}).

The spectrum is obviously flavour and chiral degenerate, and that
equation (\ref{disqcd}) holds separately for each flavour.
For simplicity let us take $\vec k$ along the z axis.  $V_{QCD}$ has a simple
meaning: up to a sign it is equal to $\partial \omega(\vec k)_{QCD}/\partial
k_z$, the group velocity of the wave packet. From eqs.
(\ref{hleft})-(\ref{aright}) it is easy to see that, at leading $T^2$ order,
$s_{QCD}=2$ and the group velocity is $\pm 1/3$.
 More precisely,
for a given value of $k_z$ and a given flavour, there are four solutions: two
possible chiralities and two possible helicities. The group velocity is given
by the eigenvalue of $\sim 1/3 \gamma_5 \sigma_z = 1/3 \gamma_5 \sigma_z (\hat
k_z)^2$ i.e. $ 1/3 \chi h \hat k_z$, where $h$ is twice the helicity, $\chi$
the chirality and $\hat k_z=k_z/\vert k_z\vert$. As a
result, the group velocity has the same sign as the momentum when the
chirality equals helicity. The solutions with opposite chirality and helicity
are disregarded at zero $T$ since they correspond to negative energy states. At
finite $T$, due to the shift of the energy, part of these branches become
physical. They are often called ``abnormal'' branches. As a consequence, for a
given value of $k_z$, a normal and an abnormal branch have opposite slopes and
there is generally one level crossing in the spectrum. It is also obvious from
the above that positive (negative) group velocity corresponds to
$\gamma_5=\sigma_z$ ($\gamma_5=-\sigma_z$). The two normal (abnormal) branches
are degenerate.

For large $\vert \vec k\vert$ the dispersion relations become non linear and
the residues of the abnormal branches are exponentially small $\sim
exp(-\vert \vec k\vert^2/\omega^2_{QCD}), \vert \vec k\vert \gg\omega_{QCD}$,
implying that abnormal excitations
 become irrelevant at large momentum\footnote{This, as
well as the doubling of the number of poles corresponding to particles
carrying a given charge, is quite natural when recalling that annihilation
operators can act non-trivially on a thermal state, to create a physical
``hole''. Clearly, for momenta so large that the corresponding state is not
thermally occupied, we expect a strong suppression.}. However, we will
restrict to the region where the linear approximation is appropriate, i.e. for
momentum $\vert \vec k\vert\ll \omega^0_{QCD}$.

In what follows, we consider the electroweak contributions that lift the chiral
and flavour degeneracy of the spectrum, together with subleading effects. We
discuss separately the asymptotic spectrum in the
unbroken and broken
phases, which differ in the quark mass effects. In the real situation, with
both phases separated by a thin wall interface, it is necessary to connect
both asymptotic spectra and further approximations will be performed.
In the following subsections, we start by considering the
toy example of one generation in order to disentangle the mixing effects and,
finally, we formulate the problem for several generations. Weak effects are not
taken into account  in the damping rate, which will then be the same
for both chiralities and every flavour:
\begin{eqnarray}
\Gamma_{L} = \Gamma_{R} = \Gamma_{QCD}.
\end{eqnarray}


\subsubsection{Asymptotic Spectrum for one generation.}
\label{sec-asun}

\begin{itemize}
\item {\bf Unbroken Phase}
\end{itemize}

The procedure is the same as in the case of QCD,  although taking
the functions $h(\omega,\vec k)$ and $a(\omega,\vec k)$ as given by eqs.
(\ref{hh}), (\ref{aa}) with vanishing masses and
$\alpha_w \ne0$, for one generation. They will be different for right and left
chiralities: the degeneracy of the normal (abnormal) branches is lifted.
We start with the one-loop self-energies in the mass basis,

\begin{eqnarray}
i\gamma_0(S^{-1})^u= \left(\begin{array}{cc} \omega + h^u_{R} -
\vec{\sigma}\cdot\vec{k} (1+a^u_{R})
- 2i \gamma
 & 0 \\ 0 & \omega + h^u_{L} + \vec{\sigma}\cdot\vec{k} (1+a^u_{L}) - 2i \gamma
\end{array}
\right),
\end{eqnarray}
where the index $u$ refers to the unbroken phase, and the functional dependence
of $h$'s and $a$'s is kept implicit. The on-shell states correspond to the
zeros of the determinant of $i\gamma_0(S^{-1})^u$, and the corresponding
eigenstates verify the effective Dirac equation
\be i\gamma_0(S^{-1})^u \psi =0.\label{direff}
\ee

At $\vec k=0$, the solutions are given by the equation
\begin{eqnarray}
{w}^u_{L,R} + {h^u}_{L,R} ({w}^u_{L,R},0)-2i\gamma = 0.
\label{poles}\end{eqnarray}

As we are interested in the low momentum quasi-particles, we expand the
effective Dirac operator around
these poles and linearise in momentum. We also neglect $O(\gamma/g_s T)$,
\begin{eqnarray}
 i\gamma_0(S^{-1})^u\frac {1\mp \gamma_5}2 = [ 1 +
\frac{\partial{h}}{{\partial{\omega}}}|_{\omega^u_{L,R}}] (\omega -{w}^u_{L,R})
-\gamma_5  (1+a^u_{L,R}) \vec{\sigma}\cdot\vec{k}, \label{su}
\end{eqnarray}
where we have used the momentum independence of $\gamma$.

The solution for the dispersion relation at small momentum is then,
\be
w(\vec{k})_{L,R}= {\omega^u}_{L,R} +\chi (\vec \sigma \cdot \vec k) V^u_{L,R}-
i \Gamma^u_{L,R}
\label{disu}\ee
where, as before, $\omega^u_{L,R}=\Re(w^u_{L,R})$ and
\be
V^u_{L,R}= \frac{1+a^u_{L,R}({\omega}^u_{L,R},0)}{
 s^u_{L,R}({\omega}^u_{L,R},0)},\qquad
\Gamma^u_{L,R}=\frac{2\gamma}{s^u_{L,R}({\omega}^u_{L,R},0)},
\label{alr2}\ee
with
\be
s^u_{L,R}({\omega}^u_{L,R},0)=1+\frac{\partial
h^u_{L,R}({\omega},0)}{\partial \omega}\vert_{{\omega}^u_{L,R}}.
\label{est2}\ee

The spectrum (\ref{disu}) is illustrated in fig. \ref{fig-sp2}(a). Even without
quark masses,
the degeneracy between L, R is now broken by the weak interactions due to
thermal effects.

\begin{figure}[t]   
    \begin{center} \setlength{\unitlength}{1truecm} \begin{picture}(5.0,3.0)
\put(-6.5,-6.0){\includegraphics{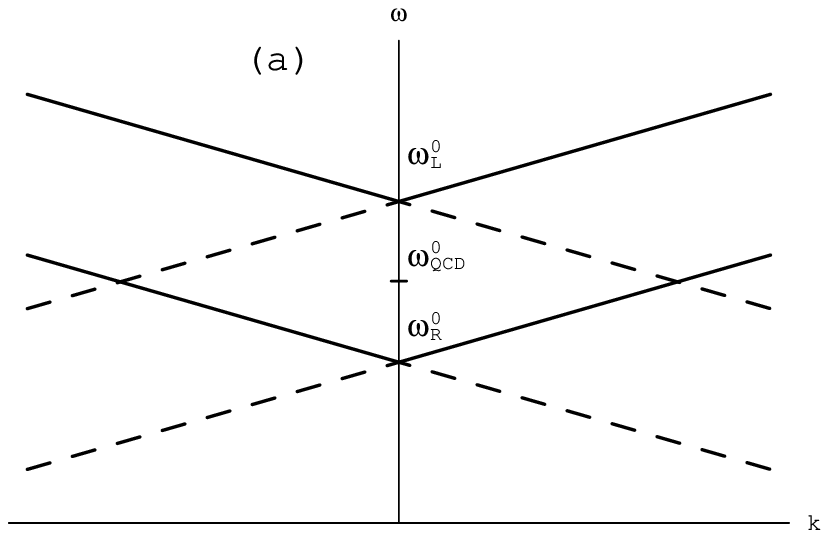}}
\put(2.5,-6.0){\includegraphics{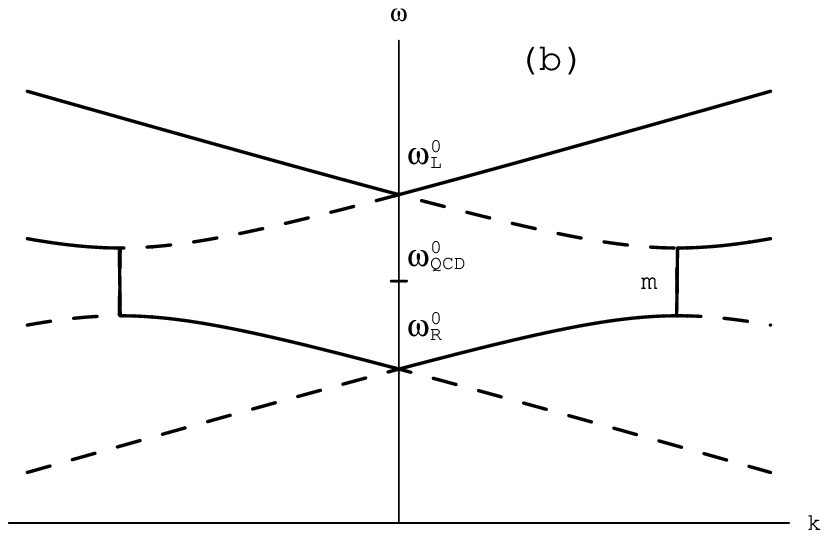}}
       \end{picture} \end{center} 	\vskip 2.6cm
\caption[]{\it{Dispersion relations for quasi-particles in the
(a) unbroken and (b) broken phases. The full (dashed) lines are normal
(abnormal) branches. The upper (lower) lines correspond to left (right)
chirality. When only QCD loops are considered, upper and lower branches are
degenerate and intersect the vertical axis at $\omega^0_{QCD}$.  The vertical
lines in (b) represent the gaps of width $\simeq m$, in which total reflection
occurs.
}} \protect\label{fig-sp2}
\end{figure}

In the regime of low momentum (\ref{disu}), the dispersion relation is well
described
by the following effective Lagrangian for free quasi-particles:

\bea {\cal L}_{eff}=&{\Psi_L}^\dagger (i \partial_t -  {i}V^u_L \vec
\partial
\cdot
\vec \sigma -{\omega}^u_L)\Psi_L +
{\Psi}_R^\dagger (i \partial_t + V^u_R{i} \vec \partial \cdot
\vec \sigma -{\omega}^u_R)\Psi_R
\nn\\  &+ i {\Psi_L}^\dagger \Gamma^u_L \Psi_L + i {\Psi}_R^\dagger
\Gamma^u_R \Psi_R,
 \label{leff}\eea
where $\Psi_L$ and $\Psi_R$ are respectively the left-handed and
right-handed fields that create quasi-particles with the given chirality.
 ${\cal L}_{eff}$ is not hermitian due to the
 damping rate.

\begin{itemize}
\item {\bf Broken Phase}
\end{itemize}

In this case, the self-energy is given by
\begin{eqnarray}
i\gamma_0(S^{-1})^b = \left(\begin{array}{cc} \omega + h^b_{R} -
\vec{\sigma}\cdot\vec{k} (1+a^b_{R})- 2i \gamma
 & (1-c) m \\ m^\dagger (1-c)^\dagger & \omega + h^b_{L} +
\vec{\sigma}\cdot\vec{k} (1+a^b_{L}) -2i \gamma \end{array}
\right).
\label{ub2}
\end{eqnarray}
The dispersion relations $w(\vec{k})$ defining  quasi-particles are
given by the zeros of the determinant of (\ref{ub2}). In the real part, the
effect of the mass is to replace the two lines of the unbroken spectrum by a
hyperbola.  This is again quite similar to the $T=0$ scenario, except that the
center of the hyperbolas is not necessarily at the origin. Between the minimum
and maximum of the hyperbolas branches, an energy gap of width $\sim m$
appears around the center. In this energy gap no plasma excitation can
exist. This new spectrum is illustrated in fig.
\ref{fig-sp2} (b).

These very narrow energy gaps are physically crucial since they induce total
reflection of a quasiparticle coming from the unbroken phase with its energy
precisely ranging in this gap. However, as we shall argue later, the large
value of the damping rate prevents the existence of quasiparticles with such a
small energy band.

In order to find the asymptotes we first consider the case $\vec k=0$ with
vanishing non-diagonal terms in eq. (\ref{ub2}). The centers of the hyperbolas
are given by
\begin{eqnarray}
\omega^b_{L,R} + {h^b}_{L,R} ({\omega}^b_{L,R},0) = 0
\label{polesb}
\end{eqnarray}
They are slightly different to those in the
unbroken phase, real part of eq. (\ref{poles}), due to mass effects in $h^b$.
The solution with $m=0$, $|\vec k|$ small, is similar to the one in the
 preceding subsection (\ref{disu}), with different values
of $V_{L,R}$ and $s_{L,R}$. A numerical estimate for down quarks gives a group
velocity
$V^b_L=V^b_R=0.339$ and $s^b_{L,R}({ \omega}^b_{L,R},0)=1.89$, close to the
values $1/3$ and $2$, respectively, obtained with the unbroken loop
at order $T^2$. For $u$ and $c$ quarks we obtain
$V^b_L=V^b_R=0.346$ and $ h^b_{L,R}({\omega}^b_{L,R},0)=1.88$, while for the
top the results are $0.165$ and $2.5$, respectively.

In the same way as we did for the unbroken phase, it is possible to obtain the
effective Lagrangian for quasi-particles in the broken phase for the low
momentum regime,

\bea {\cal L}_{eff}=&\Psi_L^\dagger (i \partial_t -  {i}V^b_L \vec
\partial
\cdot
\vec \sigma -{\omega}^b_L)\Psi_L +
\Psi_R^\dagger (i \partial_t + V^b_R{i} \vec \partial \cdot
\vec \sigma -{\omega}^b_R)\Psi_R
\nn\\  &+ i \Psi_L^\dagger \Gamma_L \Psi_L + i \Psi_R^\dagger
\Gamma_R \Psi_R
-  (\Psi_L^\dagger \mu \Psi_R + \Psi_R^\dagger \mu^\dagger
\Psi_L)
 \label{leffb}\eea
where $V^b_{L,R}, \Gamma^b_{L,R}$ are defined in terms of $h^b, a^b$
analogously to eqs.
(\ref{alr2}) and (\ref{est2}), and
\be
 \mu=\frac{m(1-c^b({{\omega}}^b_R,0))}{{s^b_{L}}^{1/2}
({{\omega}}^b_{L},0)\,{s^b_{R}}^{1/2}({{\omega}}^b_{R},0)}.\label{mu}\ee

\begin{itemize}
\item {\bf The Interface Unbroken-Broken}
\end{itemize}

The real problem we want to solve is the scattering of quasi-particles on the
interface between the broken and unbroken phases. The usual procedure in
scattering theory involves matching the in-states (asymptotic states in the
unbroken phase) to the out-states (asymptotic states in the broken phase)
through a unitary scattering matrix. The problem is, however, that the
spectrum of single quasi-particle states do not form a complete basis, or in
other words, the residues of all the single quasi-particle states do not sum
up to one{\footnote This is usually associated with collisionless
 damping which is neglected here
altogether. In massless QCD (as in NR plasmas), this type of Landau damping
only becomes relevant for large enough momenta and we expect the same to be
true here.
}. In fact, we have seen that these residues $s_{L,R}^{u,b}$ are
different for L or R quasi-particles and depending on whether they are in the
broken or unbroken phase. This is a small correction coming from the
subleading effects in T, so in order to simplify the problem we will neglect
them when obtaining the asymptotic spectrum. Thus, we will only include in the
determination of the quasi-particle dispersion relations the following
corrections:
\begin{itemize}
\item QCD corrections at leading T.
\item Diagonal Weak Interactions corrections leading in T.
\item Linear terms in momentum.
\end{itemize}
This approximation ensures that the residues of the dispersion relations (or
wave function
renormalisation constants) are the same at both sides of the wall and
equal for both chiralities.

In this approximation, the poles  in the unbroken phase are given by the
equation
\begin{eqnarray}
{w^0}_{L,R} + {\bar{h}}_{L,R} ({w^0}_{L,R},0)-2i\gamma = 0,
\label{poles2}
\end{eqnarray}
where ${\bar{h}}_{L,R}(\omega,k)$ contains only the leading-T
QCD and diagonal weak corrections of (\ref{hh}).
Now, expanding $i\gamma_0(S_0^{-1})^u$ around these poles and linearising in
momentum,
\begin{eqnarray}
i\gamma_0(S_0^{-1})^u\,\frac{1\mp \gamma_5}2 = [ 1 +
\frac{\partial{\bar{h}}}{{\partial{\omega}}}|_{{\omega^0_{L,R}}}] (\omega
-{{w^0}}_{L,R}) -\gamma_5  (1+\bar{a}^u_{L,R}) \vec{\sigma}\cdot\vec{k},
\label{s0u}
\end{eqnarray}
with ${\bar{a}}_{L,R}(\omega,k)$ containing the leading-T, QCD and diagonal
weak corrections in the functions (\ref{aa}).
Using eqs.(\ref{hleft})-(\ref{aright}), together with the expansion $F(x)\sim 1
+\frac{1}{3 x^2},
x\gg 1$ we get, to leading order in $\gamma/(g_sT)$,
\begin{eqnarray}
1+ {\bar{a}}_{L,R}({\omega^0}_{L,R},0) = \frac{2}{3},\label{unpa}
\end{eqnarray}
\begin{eqnarray}
1 + \frac{\partial{\bar{h}_{L,R}}}{\partial{\omega}}|_{{\omega^0}_{L,R}} = 2,
\label{undh}\end{eqnarray}

Notice that although $\bar{h}_{L,R}$ is flavour dependent, the pole condition
in eq. (\ref{poles2}) implies that the above results, eqs. (\ref{unpa}) and
(\ref{undh}) are not. The damping rate is
\be
\Gamma = \frac{2\gamma}{1+\frac{\partial{\bar{h}_{L,R}}}
  {\partial{\omega}}\vert_{{\omega^0}_{L,R}}} = \gamma.
\ee
The dispersion relations describing the quasi-particles
in the unbroken
phase are given by the zeros of eq. (\ref{s0u}):
\begin{eqnarray}
w(k)_{L,R} = {\omega^0}_{L,R} +\chi \frac{1}{3} \vec{\sigma}\cdot\vec{k} - i
\gamma.
\label{ayyy} \end{eqnarray}
In the broken phase, the only difference with respect to eq.(\ref{ayyy}), in
this approximation, results from the $L, R$ mixing
effects due to the tree level mass. Thus, using eq. (\ref{s0u}),

\begin{eqnarray}
i\gamma_0(S_0^{-1})^b =i\gamma_0(S_0^{-1})^u + \left(\begin{array} {cc} 0
 &  \frac{m}{2} \\ \frac{m^\dagger}{2} & 0 \end{array}\right),
\label{s0b}
\end{eqnarray}
where we have neglected the functions $c(\omega,k)$ because they are subleading
in T. The dispersion relations in the broken phase are just given by the
determinant of $i\gamma_0(S_0^{-1})^b$.

{}From (\ref{s0u}) and (\ref{s0b}), we can easily write the effective
unperturbed Hamiltonian for quasi-particles
in the presence of a wall, in the basis of asymptotic states in the unbroken
phase,

\be
H^0_{eff}=\left(\begin{array}{cc
} -\frac{1}{3} i \sigma_z\partial_z   -i \gamma  + {{\omega}}^0_R & \frac{m}{2}
\theta(z) \\ \frac{m}{2} \theta(z) & \frac{1}{3} i \sigma_z\partial_z  -i
\gamma +{{\omega}}^0_L \end{array}
\right).
\label{heffo}
\ee

\subsubsection{Several generations.}
\label{sec-several}

In the case of several generations, the exact solution of the spectrum
is very complicated, as the functions
(\ref{aa}), (\ref{hh}), (\ref{ac}) are non-diagonal $3 \times 3$ matrices in
flavour space.
In particular, $h(\omega, k)$ and $\,\,1+a(\omega,k)\,\,$ may not be
diagonalisable in the same basis, in such a way that single
quasi-particle asymptotic states cannot be chosen to be diagonal in flavour.
In order to simplify
the problem, we will derive the dispersion relations for asymptotic
quasi-particle states neglecting flavour mixing
and, as we did for the one-flavour case, neglecting also the subleading effects
in T,
both in QCD and weak corrections.
As previously stated, in this way the residues
of the dispersion relations
are the same at both sides of the wall, for any flavour and chirality.

We start with the $6\times 6$ one-loop effective Dirac operator in the mass
basis:

\begin{eqnarray}
i\gamma_0(S^{-1})^u = \left(\begin{array}{cc} \omega + h^u_{R} -
\vec{\sigma}\cdot\vec{k} (1+a^u_{R})
- 2i \gamma
 & 0 \\ 0 & \omega + h^u_{L} + \vec{\sigma}\cdot\vec{k} (1+a^u_{L}) -2i \gamma
\end{array}
\right),
\end{eqnarray}
\begin{eqnarray}
i\gamma_0(S_0^{-1})^b = \left(\begin{array}{cc} \omega + h^b_{R} -
\vec{\sigma}\cdot\vec{k} (1+a^b_{R})- 2i \gamma
 & (1-c) m \\ m^\dagger (1-c)^\dagger & \omega + h^b_{L} +
\vec{\sigma}\cdot\vec{k} (1+a^b_{L}) -2i \gamma \end{array}
\right).
\label{ub}
\end{eqnarray}

The results of the previous section hold here, if we substitute
all $\omega^0_{L,R}$ and $h$, $a$ and $c$ by $3\times 3$ matrices. The
expressions for $h^b$,  $a^b$ and $c^b$ ($h^u$,  $a^u$ and $c^u$) were given in
subsection (\ref{sec-realun}) with masses different from (equal to) zero. The
effective
unperturbed Hamiltonian for free quasi-particles is then
\be
H^0_{eff}=\left(\begin{array}{cc
} -\frac{1}{3} i \sigma_z\partial_z   -i \gamma  + {\hat{\omega}}^0_R &
\frac{m}{2} \theta(z) \\ \frac{m}{2} \theta(z) & \frac{1}{3} i
\sigma_z\partial_z  -i \gamma +{\hat{\omega}}^0_L \end{array}
\right),
\label{heffo2}
\ee
where ${\hat{\omega}}^0_R$ and ${\hat{\omega}}^0_L$ are $3\times 3$ diagonal
matrices $({\omega^0}^1_R, {\omega^0}^2_R, {\omega^0}^3_R)$,  $({\omega^0}^1_L,
{\omega^0}^2_L, {\omega^0}^3_L)$, respectively.

 In section (\ref{sec-CP}), we will see that it is possible and consistent to
include the remaining effects (non-leading and mixing) as a perturbation,
without modifying the asymptotic states.

\subsection{The damping rate.}
\label{sec-damping}

Due to incoherent thermal
scattering with the medium, the energy and momentum of the quasi-quarks in a
plasma are not
sharply defined, but spread like a resonance of width $\gamma$
\cite{bp}.

The
QCD damping rate at zero momentum is of
the order $\gamma \sim 0.15\, g_s^2\, T$ \cite{bp}, i.e. $\sim 19$ GeV at
$T=100$ GeV. Compared to the  energy of the quasiparticle, of $O(g_sT)$, the
damping rate is formally small, i.e. it contains one additional power of $g_s$.
This hierarchy allows to speak of the mere existence of quasiparticles as
coherent excitations \footnote{In practice however $g_s$ is not so small, and
the width of $\sim 20$ GeV is not really small compared to the energy $\sim
50$ GeV. We could say that quasi particles hardly exist. The plasma is mainly
an incoherent mixture of states.}.

The quasi-particle has thus a finite life-time $\sim 1/2\gamma$, turning
eventually into a new state, out of phase with the initial one. As the group
velocity is of order $1/3$, the mean free path is of order $1/6 \gamma\sim
1/120$ GeV$^{-1}$.  Even if it is true that the relative phase between
different quark flavours is not destroyed by collisions with the QCD thermal
bath, it should be stressed that the quantum spatial coherence, i.e. the phase
relation between points separated by a distance $\ge 1/6\gamma$, is lost.

The imaginary part of the QCD self-energy is  much
larger than the real part of the electroweak self-energy. For all quarks but
the top, it is also much larger than the corresponding mass gap in the broken
phase discussed in subsection (\ref{sec-asun}). In other words, as we shall see
explicitly later, the coherent electroweak processes relevant in the following
require a time much larger than the mean free time of the quasiparticles.

\seCtion{Reflection on the bubble wall: one flavour case.}
\label{sec-wall}

 In this section we discuss a world with just one flavour, and the tree-level
reflection properties when a quasi-particle hits the boundary (thin wall
approximation) separating the two phases of spontaneous symmetry breaking. No
electroweak effect is considered, other than the mass of the quasi-particle. We
quantitatively derive the dramatic effect of the QCD damping rate on the
reflected density of quasi-quarks. After some considerations in terms of damped
waves, the system is described in terms of wave-packets which exemplify the
mean free path and mean free time of the states. We discuss as well the
associated density matrix formalism.


\subsection{Waves and wave packets}

  Consider the effective hamiltonian, eq.(\ref{heffo}). As stated above,
electroweak loops are irrelevant for the purpose of this section, in which case
we could settle $\omega_L^0\,=\,\omega_R^0\,=\,\omega_{QCD}^0$ (see
eq.(\ref{omega0})). The $L,R$ dichotomy will be relevant in the next section.
The effective Dirac equation describing the quasi-particle interaction with the
wall is:

\be
i \partial_t \psi= H_{eff}\,\psi = \left(\begin{array}{cc
} -\frac{1}{3} i \sigma_z\partial_z   -i \gamma  + {{\omega}}^0_{QCD} &
\frac{m}{2} \theta(z) \\ \frac{m}{2} \theta(z) & \frac{1}{3} i
\sigma_z\partial_z  -i \gamma +{{\omega}}^0_{QCD} \end{array}
\right)\psi.
\label{direff2}
\ee
 There are many solutions to this differential equation. The problem to solve
is the scattering of a quasi-quark hitting the wall from the unbroken phase.
The initial conditions correspond then to a state localised on $z<0$ at its
creation time, travelling towards the $z>0$ region, and with a mean free time
and mean free path as described in subsection \ref{sec-damping}.

 No single wave solution fulfills all these conditions. The gist of the
problem is reflection, though, and  waves damped in space and travelling
towards the wall should be an appropriate heuristic treatment of the initial
state, that we proceed to develop here. These solutions have then a mean free
path (coherence length), but are eternal,
that is, have no mean free time. All the initial conditions can be accounted
for in term of wave packets instead,
and we will see later that this  physically more correct approach leads to the
same conclusions.

  An exponentially decaying wave, created in the unbroken phase at a 	point
$z_0<0$ within a mean free path ($1/6\gamma$) away from the wall, and
travelling towards $z>0$, is given by

\bea
\psi^\chi_{inc}(z,t;\omega)=e^{-i \omega t} & \theta(z-z_0) \bigg\{ &
  \theta(-z)\left[e^{ i\pi_\chi    z}                u_{ \chi,\chi}+
                  e^{-i\pi_{-\chi}z} r_\chi(\pi_\chi) u_{-\chi,\chi}
            \right]\nn\\
 & + &\theta(z) e^{i\pi^t_\chi z}
   [u_{\chi,\chi} + r_\X(\pi_\chi) u_{-\chi,\chi}]\bigg\},
\label{planew}
\eea

 where $\chi$ denotes chirality and  $u_{\X,\sigma}$
is the normalised Dirac spinor for a particle with spin $\sigma_z/2$ and
chirality $\X$. $\pi_\chi$ is the incoming momentum of a particle with
$\sigma_z=\chi$, i.e.,
positive group velocity

\be
\pi_\chi= 3 (\omega-\omega^0_\X+i\g)= p_\X +3 i\g,\qquad \pi_{-\chi}=
3 (\omega-\omega^0_{-\X}+i\g),
\label{pichi}\ee
 where $p_\X=\Re(\pi_\X)$. $\pi_\chi^t$ is the corresponding  transmitted
momentum  in the broken phase,
 \be
\pi_\chi^t={1\over2}(\pi_\chi-\pi_{-\chi})
  +{1\over2}\sqrt{(\pi_\chi+\pi_{-\chi})^2-9\,m^2}.
\ee
 The reflection coefficient $r_\X(\omega)$ has the simple
expression:

\be
r_\X(p_\X)\equiv r_\X(\omega)=-\frac {m/2} {p(\omega)+
e^{i\phi}\sqrt{\vert p(\omega)^2-m^2/4\vert}},
  \label{refcoeff}\ee
with\footnote{To simplify the notations we use the same symbol $r_\X$ for the
reflection coefficient as a function of $\omega$ and as a function of $p_\X$.
The analytic behaviour of eq.(\ref{refcoeff}) will be discussed later, see
subsection \ref{sec-refprob}.}

\be \phi=\frac {\arg (p(\omega)-m/2) +\arg(p(\omega)+m/2)} 2
\label{phi}\ee
and

\be
p(\omega)
=\frac{1}{6}\,[p_\chi+p_{-\chi}]=\omega-\frac{\omega^0_\chi+\omega^0_{-\chi}}2
\,,\label{defpofw}
\ee
where we
introduced the momentum-like variable $p(\omega)$ to stress the analogy with
the situation at  $T=0$ \cite{nous0}. The analytic continuation of $r(p_\X)$ to
the complex plane for the variable $p_\X$($\omega$) is straightforward from eq.
(\ref{refcoeff}).

For quasi-quarks with current quark masses smaller than $2\gamma$,
\be
|r_\X(\pi_\X)|\equiv |r_\X(\omega+i\g)|\le {m\over 4 \g}.\label{borne}
\ee
   It follows that the reflection probability is suppressed for all flavours
but
the top. For the latter, $|r(\omega)|\le 1$, as expected.

\subsubsection{ Wave packets}
\label{sec-packets}

Our main hypothesis results from the uncertainty principle. The finite
life-time (and finite mean free path) of the quasiparticles induces a finite
spreading of their energy (and momentum) spectrum of the order of $2\gamma$
($6\gamma$). Wave packets allow a parametrisation of such a  localisation in
time and space.
We use  gaussian
wave packets with a unique
width $1/d$. Our final result will turn out to be
independent of this particular choice of packets, the only requirement being
$d\, \le\,1/3\gamma$.

Wave packet solutions to eq.(\ref{direff2}) can be expressed as superposition
of plane waves.
 Assume that, at time $t_0$, an incoming quasiparticle wave packet has been
created around the position $z_0<0$ in the unbroken phase, with a central
momentum $p_0$:

\be \Psi_{inc}^\chi(z,t_0;z_0,t_0,p_0,d,\chi)= \sqrt{\frac d
{2\pi^{3/2}}}\int_{-\infty}^{+\infty} dp
e^{-d^2(p-p_0)^2/2} e^{i p(z-z_0)}u_{\chi, \chi} \label{onde}.\ee

 The choice of spinor with $s_z=\chi$ ensures the positive group velocity,
characteristic of an incoming wave. Notice that the eigenspinors of the Dirac
Hamiltonian with positive group velocity are the same all over one branch of
the spectrum (see fig. \ref{fig-sp2} (a)), as long as we stay in the unbroken
phase and in the linear regime for the spectrum.

{}From eq. (\ref{direff2}) it is easy to derive the time evolution of the wave
function (\ref{onde}), in the unbroken phase:
\bea
\lefteqn{\Psi_{inc}^\chi(z,t;z_0,t_0,p_0,d,\chi)= \nn}\\  & &
\theta(t-t_0)\sqrt{\frac d {2\pi^{3/2}}}\int_{-\infty}^{+\infty} dp_\X
e^{-d^2(p_\X-p_0)^2/2} e^{i p_\X (z-z_0)} e^{-i (\omega-i\gamma) (t-t_0)}
u_{\chi, \chi} \nn\\  & &
=\theta(t-t_0)\sqrt{\frac 1 {d\pi^{1/2}}}e^{-\frac
{[(z-z_0)-1/3(t-t_0)]^2}{2d^2}}e^{i p_0
[(z-z_0)-1/3(t-t_0)]}e^{-i\omega^0_\chi(t-t_0)-\gamma (t-t_0)} u_{\X,\X},
\label{wps}\eea
where the relation between the incoming momentum, $p_\X$, and $\omega$ was
given in eq. (\ref{pichi}). Eq (\ref{wps}) exhibits the motion of the center of
the packet with velocity 1/3.
The Fourier transform of eq.(\ref{wps}) is
\bea \lefteqn{\widetilde \Psi_{inc}(p_\chi,t;z_0,t_0,p_0,d,\chi)= \nn}\\ & &
\theta(t-t_0)\sqrt{\frac d {\pi^{1/2}}}
e^{-d^2(p_\chi-p_0)^2/2} e^{-ip_\chi z_0} e^{-i(\omega-i\gamma) (t-t_0)}
u_{\chi, \chi}. \label{wpf}\eea

The wave function is normalised to 1 at $t=t_0$, and its norms decays
exponentially as $\sim e^{-2\g (t-t_0)}$ due to dissipation. Let us now turn to
a statistical ensemble of particles. Consider, in the unbroken phase, an
homogeneous flux of such wave packets  travelling towards
 the wall. These particles ``decay'', i.e. they are
scattered, at a rate $2\g$ during their flight.
This decay must be exactly cancelled by  creation at the
same rate $2\g$, so that the equilibrium distribution of particles can
remain stationary. To account for this effect,
we introduce  a continuous differential rate of creation $N(z_0,t_0,p_0) dz_0
dt_0 dp_0$.
The density  of incoming quasi-particles of chirality and spin $ \X$ at some
point $z<0$ {\it far from the wall} and at time $t$ is given by

\bea
\Xi&=\int_{-\infty}^{+\infty} dp_0 \int_{-\infty}^{t}dt_0
\int_{-\infty}^{+\infty}dz_0 \vert
\Psi_{inc}^\chi(z,t;z_0,t_0,p_0,d)\vert^2N(z_0,t_0,p_0) \nn\\ &=
\frac 1 {2\gamma} \int_{-\infty}^{+\infty} dp_0
N(z_0,t_0,p_0).\label{Ndef}\eea

Taking for the rate of quasi-particle creation
\be N(z_0,t_0,p_0) =  2  \gamma n_F(\omega_\chi^0 + p_0/3, T),\label{Ndeux}\ee
 the equilibrium Fermi density is recovered,
\be \Xi= \int dp_0 n_F(\omega_\chi^0 + p_0/3, T), \label{densit}\ee
where $n_F(\omega)=1/(\exp(\omega/T)+1)$.

\subsubsection{Reflection probability.}
\label{sec-refprob}

The reflected part of the wave packet, travelling backwards in the unbroken
phase is

$$
\Psi_{ref}^\chi(z,t;z_0,t_0,p_0,d)= 
\theta(t-t_0)\sqrt{\frac d {2\pi^{3/2}}}\int_{-\infty}^{+\infty} dp_\X$$\be
e^{-d^2(p_\X-p_0)^2/2} e^{-i p_{-\X}z-i p_\X z_0} e^{-i (\omega-i\gamma)
(t-t_0)}
r_\chi(p_\X)
u_{-\chi, \chi}, \label{wpr}\ee
where the $\omega$ dependence is the same than for $\Psi_{inc}$, as energy is
conserved in the interactions with the wall, and where $r_\chi(p_\X)$ has been
defined in eq. (\ref{refcoeff}). In eq. (\ref{wpr}), $\chi$ ($-\chi$)represents
the incoming (outgoing) chirality.

Let us define the reflection probability $n_r$ as the ratio of the reflected
flux close to the wall on the unbroken side ($z=0^-$), to the incoming flux.
Since the incoming and outgoing group velocities have the same absolute
values, it amounts to the ratio between the reflected density at $z=0^-$ and
the incoming density. Notice that the choice ($z=0^-$) maximises $n_r$, as the
reflected flux will be diluted when travelling backwards in the unbroken
phase.  Further than a few mean free paths away from the wall, the reflected
flux of particles is transported by {\em diffusion} rather than {\em free
streaming}. As we will see in the next section,
diffusion could be qualitatively described by replacing (\ref{Ndeux}) with its
local equilibrium generalisation. For simplicity, we concentrate now
on the reflected flux close to the wall.

At time $t=0$, the reflection probability is obtained integrating the
reflected density, eq. (\ref{wpr}), over all creation times in the past,
$t_0<0$, and over all creation locations in the unbroken phase, $z_0<0$,
\be n^\X_r(z=0,t=0)= \frac 1 {\Xi}\int_{-\infty}^{+\infty} dp_0
\int_{-\infty}^{0} dz_0 \int_{-\infty}^{0} dt_0  N(z_0,t_0,p_0)
 \left\vert \Psi^\X_{ref}(0,0;z_0,t_0,p_0,d,\chi) \right\vert^2.
\label{refprob}\ee
This is consistent with the creation probability  eq. (\ref{Ndef}).
$N(z_0,t_0,p_0)$ was given in eq. (\ref{Ndeux}).

The computation of  the reflection probability, eq. (\ref{refprob}),
requires a careful analysis of the {\it mathematical} properties of the
function $r_\X$, eq. (\ref{refcoeff}).
$r_\X$ is analytic on the complex plane but for a cut extending on the real
axis between $p(\omega)=-m/2$ and $p(\omega)=m/2$. Close to the cut, $\phi
\simeq \pi/2$ in the upper half-plane,  while $\phi \simeq -\pi/2$ in the lower
half-plane. When integrating on the real axis we will take the determination
from the upper half-plane.  Outside the cut on the real axis and for
$p(\omega)>m/2$, $\phi=0$ while for $p(\omega)<-m/2$, $\phi=\pi$\footnote{Then,
on the real axis, $r_\X =m/2/(p(\omega)+\sqrt{p(\omega)^2-m^2/4})$ for
$p(\omega)>m/2$, and $r_\X =m/2/(p(\omega)-\sqrt{p(\omega)^2-m^2/4})$ for
$p(\omega)<-m/2$.}. It results that $\vert r_\X(\omega) \vert \le 1$. More
precisely, $\vert r_\X(\omega) \vert = 1$ for $-m/2< p(\omega)<m/2$, and
$\vert r_\X (\omega) \vert < 1$ outside that segment. This is illustrated in
fig. (\ref{fig-r}).  Fig (\ref{fig-phi}) displays the phase of
$r_\chi(\omega)$, and shows that it varies very fast over the real axis on the
cut. It is worth to notice that this cut does correspond to the total
reflection domain for plane waves. It is a very small domain for light quarks.

\begin{figure}[t]   
    \begin{center} \setlength{\unitlength}{1truecm} \begin{picture}(6.0,6.0)
\put(-6.0,-9.0){\includegraphics{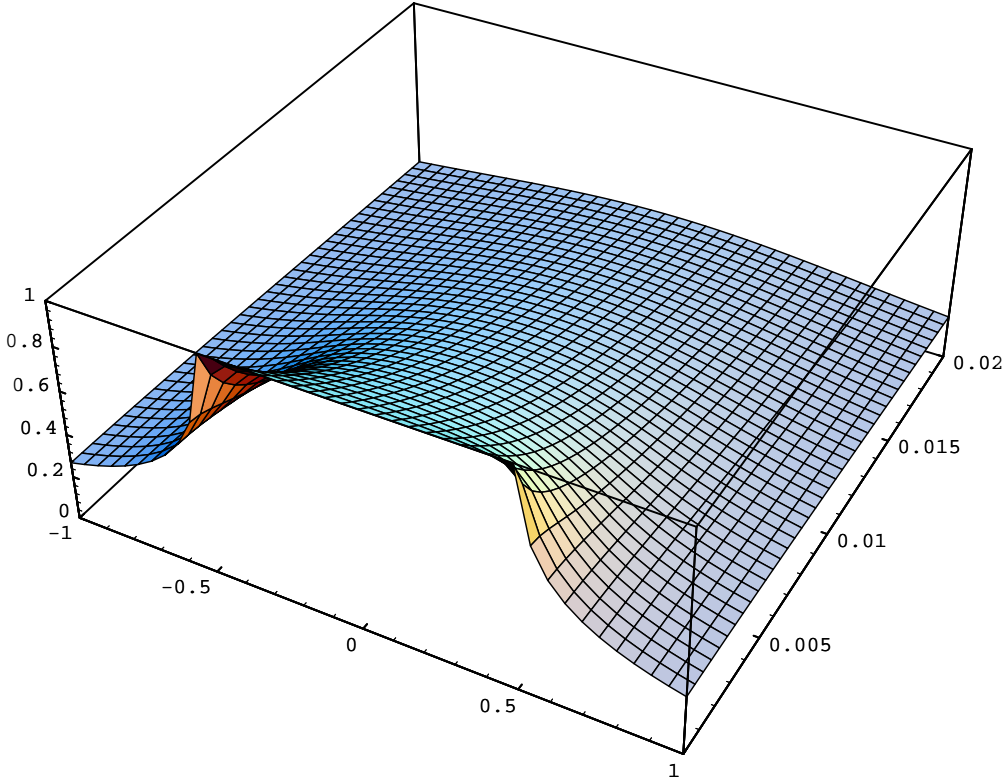}}
       \end{picture} \end{center} 	\vskip 2.6cm
\caption[]{\it{Plot of
$|r(p)|$ on the upper complex $p(\omega)$ half-plane, with
 $m = 0.5$.  The lower left edge is the real axis.
The crest at $|r(p)|=1$ on this axis corresponds to total reflection. Notice
the rapid decrease of $|r(p)|$ with increasing imaginary part.}}
\protect\label{fig-r}
\end{figure}

\begin{figure}[t]   
    \begin{center} \setlength{\unitlength}{1truecm} \begin{picture}(6.0,6.0)
\put(-6.0,-9.0){\includegraphics{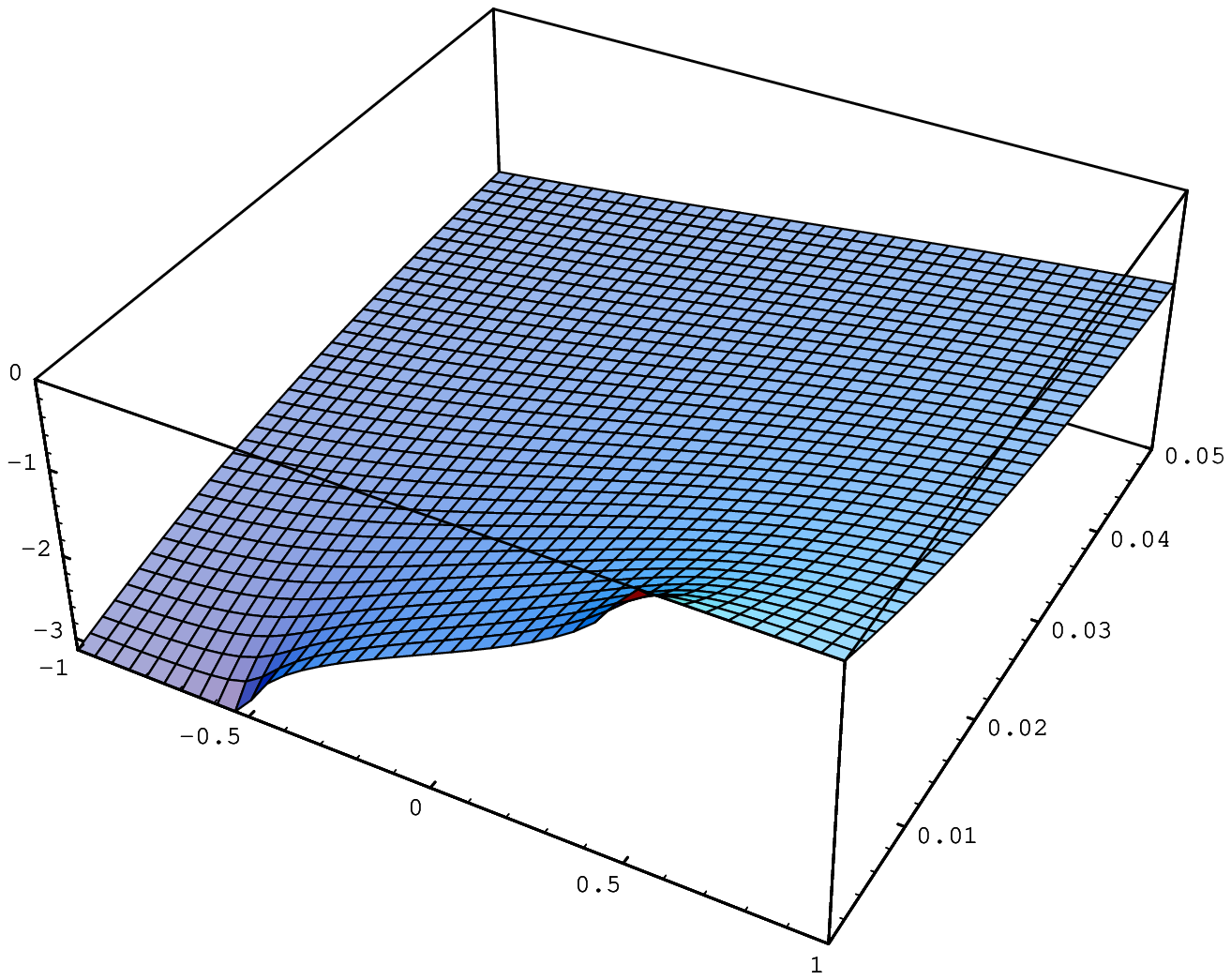}}
       \end{picture} \end{center} 	\vskip 2.6cm
\caption[]{\it{Plot of the
phase of  $r(p)$ on the upper complex $p(\omega)$ half-plane, with $m = 0.5$.
The phase is zero on the real axis outside the total reflection domain,
$-0.5<x<0.5$, and varies rapidly inside this domain. This explains the
suppression observed when the integration is performed on the real axis.}}
\protect\label{fig-phi}
\end{figure}

In eq. (\ref{wpr}) the integral over $p_\X$ is strongly reduced by the rapid
rotation of the phases of the exponential and the function $r_\X$. This is at
the origin of the dramatic suppression of CP asymmetry that we advocate in this
paper.
We have checked this suppression numerically, although the same result follows
from a
very simple analytic argument, which we proceed to develop.
 The only singularity in the integral $dp_\X$ in eq. (\ref{wpr}) is the cut on
the real axis of the function $r_\X$. Consider a rectangle in the complex
$p_\X$ plane with sides given by : the real axis, a parallel to it in the upper
half-plane, and two vertical sides at $\pm \infty$. The extension of the
integral in eq. (\ref{wpr}) to this contour vanishes since no singularity is
present inside the box. The integrals on the vertical sides vanish due to
gaussian suppression. The integral in eq. (\ref{wpr}) on the real axis can then
be replaced  by an integration on a parallel axis, $\int_{-\infty +3 i \beta
\gamma}
^{\infty +3 i \beta \gamma} dp_\X$, with $0 \le \beta \le 1$. The  condition
$\beta\ge 0$ ensures that the integral over $z_0$ in (\ref{refprob}) is finite,
while the condition $\beta\le 1$ does the same for the integral over $t_0$.

The choice $\beta=1$, and thus $\Im(p_\X)= 3 i \gamma$
is equivalent to change variables $p_\X \ra p_\X +3 i \gamma$ in eq.
(\ref{wpr})\footnote{ Notice that $p_\X +3 i \gamma$ is the variable $\pi_\X$
defined in eq. (\ref{pichi}).}:

\bea
\lefteqn{\Psi_{ref}^\chi(z,t;z_0,t_0,p_0,d)= \nn}\\  & &
\theta(t-t_0)\sqrt{\frac d {2\pi^{3/2}}}\int_{-\infty}^{+\infty} dp_\X
e^{-d^2(\pi_\X-p_0)^2/2} e^{-i \pi_{-\X}z-i \pi_\X z_0} e^{-i \omega (t-t_0)}
r_\chi(\pi_\X)
u_{-\chi, \chi}. \label{wprcomp}\eea

The reflection coefficient $r_\X(\pi_\X)$ is then bounded as displayed in eq.
(\ref{borne}). The exponential in (\ref{wpr}) also provides a factor $\exp(9
d^2 \g^2/2)$, and a bound for $\Psi_{ref}^\chi$ results as long as $d\ll
1/\g$\footnote{Physically, this inequality means that the layer of particles
sitting ``on'' the wall has
a negligible thickness $d$, compared with the layer of particles that ever
have a chance to reach the wall before being damped.
},

\be
\left\vert \Psi_{ref}^\chi(z,t;z_0,t_0,p_0,d)\right\vert \le \frac m
{4\sqrt{p(\omega_\X^0+p_0/3)^2+\gamma^2+O(m^2)}}\label{borne2} + O(3 d \g),
\ee
where $p(\omega)$ has been defined in (\ref{defpofw}) and the bound in
(\ref{borne}) has been slightly improved for later use.

{}From eq. (\ref{refprob}) it follows that
\be
n_r(0,0)\le \frac {m^2} {16 \sqrt{\g^2 +O(m^2)}\, \Xi}\label{supr}
\ee
 It is worth to notice that the results expected for zero damping rate $\g=0$
can be recovered. In this case, eq. (\ref{supr}) reduces to
\be n_r(0,0) \sim \frac m \Xi , \label{would}\ee

This is indeed the result that would follow directly for vanishing damping
rate,
when the reflection probability can be directly computed in terms of plane
waves:
\be
\frac 1 \Xi \int d p_\X n_F(\omega,T) \vert r_\chi(\omega)\vert^2\sim \frac m
\Xi \label{would2}\ee
where $\omega$ and $p_\X$ are related as in eq. (\ref{defpofw}).

The comparison of eqs (\ref{supr}) and (\ref{would}) shows a $\sim
m/(16\gamma)$ suppression of the reflection probability when the damping rate
is taken into account,
 due to the fast rotation of the phases in the integral in eq. (\ref{wpr}).
This is a typical quantum effect: the
different frequencies in the integral, eq. (\ref{wpr}),  add up as waves, not
as probabilities.
 We may express the phenomenon at work as follows: the damping rate destroys
the quantum coherence which is necessary to have a large reflection
probability, through an average over a rapidly rotating phase,

\be
<\,r_\X\,>\,<\,r^\dagger _\X\,>\,\ll\,<\,{\vert r_\X \vert}^2\,>.
\label{average}\ee

The spread in energy and momentum of the quasi-particles results in a smearing
of the reflection integrals, contrary to the case of vanishing damping rate,
where the states are monochromatic, and thus well represented by plane waves.

\subsubsection{A simple approximation to the reflection probability.}

A useful simplification is obtained if the integration on $t_0$ in eq.
(\ref{refprob}) is extended  to $+\infty$.  A qualitative estimate of the error
introduced by this trick is given here. A rigorous proof is developed in
appendix \ref{app-proof}.

 The presence of the potentially dangerous exponential  $e^{\g t_0}$ in the
expression for $\Psi_{ref}$, eq. (\ref{wpr}), does not lead to any divergence,
due to the gaussian smearing. Indeed, the only effect of extending the $t_0$
integration from $0$ to $+\infty$ is to add a gaussian tail. The width of the
gaussian being $d$, the added tail will be $O(\exp(-1/(\g d)^2)$, i.e. small as
long as $d \g \ll 1$. More precisely, as explained in appendix \ref{app-proof},
the correction is negligible for
\be
 3d \g \ll 1.
\label{condi}\ee
 The reflection probability (\ref{refprob}) is then given by

$$ n^\X _r (0,0)= \frac {2 \g}{ \Xi}\int_{-\infty}^{+\infty} dp_0
\int_{-\infty}^{0} dz_0 \int_{-\infty}^{\infty} dt_0  n_F{(\omega_\chi^0 +
p_0/3, T)} \frac d {2\pi^{3/2}}$$
\be
\int_{-\infty}^{\infty} dp_\X \int_{-\infty}^{\infty} dp'_\X
e^{-d^2(\pi_\X-p_0)^2/2-d^2(\pi'^\ast_\X-p_0)^2/2}e^{i \pi'^\ast_{\X}z_0-i
\pi_\X z_0}
e^{i (\omega-\omega') t_0}r_\chi(\pi_\X)r^*_\chi(\pi'_\X).\label{nrcomp}
\ee

The integral over $t_0$ simply gives
$2\pi\delta(\omega-\omega')=6\pi\delta(p_\X-p'_\X)$. The  $z_0$ dependence is
then  $\exp(-6 \g z_0)$, whose integral is trivial. Finally,
\be n^\X _r(0,0)= \frac 1{\Xi}\int_{-\infty}^{+\infty} dp_0  n_F{(\omega_\chi^0
+ p_0/3, T)}
\frac d {\sqrt{\pi}}\int_{-\infty}^{\infty} dp_\X e^{-d^2(p_\X-p_0)^2
+(3d\g)^2}\vert r_\chi(\pi_\X)\vert^2. \label{nrsimpl}
\ee

This is the main result of this section. After neglecting the $t_0>0$ terms and
{\it upon the change of variables $p_\X \ra p_\X + 3 i \g$ and correspondingly
$\omega \ra \omega+i\g$}, the reflected flux for wave packets is just a
gaussian smear-out of
the reflection fluxes for plane waves, as naively expected.

Two comments are appropriate. The first one concerns particle number, which
must
be conserved. Unitarity implies that the damping of reflected flux is exactly
compensated by the enhancement of the transmitted one.
 The underlying physics  can also be
understood. A very localised particle penetrates the broken phase in a time
$1/3d$, whereas for small masses it requires a rather long time $2/m$ to be
significantly
reflected (the reflection dominantly happens in an energy window of gap $\sim
m$, the total reflection domain, and by uncertainty principle this corresponds
to a time $\sim 1/m$). If in between the particle is ``measured'' in the
broken phase by collisions with the plasma, the wave function
collapses into one corresponding to a localised particle in the broken phase,
and as a
result, the particle has been effectively transmitted.

Secondly, the $d$ dependence of (\ref{wpr}) can be clarified.  When
the mass is large enough, $\g\ll 1/(3d) \ll m/2$, reflection typically occurs
between 2 collisions, and the effect of both the $\g$-shift and the
$1/(3d)$-smearing can be neglected, nicely recovering the usual
asymptotic reflection probability for plane waves.  For the opposite limit of
small masses, the $d$ dependence can also be mild. Indeed, exchanging the order
of integration in (\ref{nrsimpl}) leaves us with a thermal average of the
reflection probabilities $|r_\X(\pi_\X)|^2$, according to an effective thermal
distribution
\be
n_F^d(\w_\X^0+p_\X/3,T)={d\over\sqrt{\pi}}\int_{-\infty}^\infty dp_0
e^{-d^2(p_\X-p_0)^2 +(3d\g)^2} n_F(\omega_\chi^0 + p_0/3, T).
\ee
If $n_F$ is smooth on scales
$1/(3d)$ (as is the case, close to equilibrium, if $1/(3d)\ll T$), $n_F^d\sim
n_F$ and the average (\ref{nrsimpl}) is close to a thermal average of
monochromatic reflection probabilities.

We have  checked numerically this $d$ independence of our result. If
 the distribution probability was arbitrary and far from equilibrium (i.e. far
from $n_F(\omega)$), our
 linear response approach to the relaxation processes loses its
meaning, and finer knowledge of all relaxation processes becomes necessary.
Let us thus stress that we assumed the quasi-particle distributions to be close
to equilibrium (and
effectively reducing the reflection coefficient makes this assumption easier
to meet, for a given wall speed), and moreover required
\be
\g\ll{\frac 1 {3d}}\ll T \Leftrightarrow g_s^2 T\ll \frac 1 {3d} \ll T,
\label{cond}\ee
which is always satisfied for weak coupling $g_s\ll 1$.  Our model for
decoherence accounts for the physical consequences of a non-vanishing damping
rate. In real QCD $g_s$ is not small, and we believe that a more realistic
treatment would lead to a further suppression of the CP-asymmetry than
advocated in this paper. Indeed, as the damping rate increases the description
of the system in terms of quasi-particle ( an essential ingredient of the
approach) looses its physical meaning.

\subsection{Density matrix.}

In this subsection,   the results of the preceding one are reproduced using the
formalism of density matrices.
 The language of density matrices  is very convenient for plasma physics. It
will clarify how the suppression effect found above depends on the quantum
aspects of the density matrix, and their relation with the breaking of thermal
equilibrium in the process under study.

The effective Hamiltonian in the plasma rest frame, eq. (\ref{direff2}), can be
split into hermitian and non-hermitian
parts:
\be H_{eff} \equiv H_{herm} - i \gamma. \ee
We consider the
reduced density matrix $\rho$ for one quasi-particle at equilibrium in the
unbroken phase.  It obeys the following Boltzman
equation,

\be \partial_t \rho(t)=-i[H_{herm}, \rho(t)] + G - L\label{vn}.\ee

$G$ ($L$) is a gain (loss) term corresponding to the creation (disappearance)
of  quasiparticles due to collisions with the medium. To simplify the problem,
we used a linearised equation. In this approximation, the $L$ term is related
to the damping rate by $L=-2\g \rho$.
At equilibrium, the probability loss due to $L$ must be compensated by the $G$
term. However, and this is our essential assumption, $G$ creates
quasi-particles from the medium which are out of phase with respect to the ones
destroyed by $L$. We further assume $G$ to be homogeneous in space and time in
the unbroken phase. A simple example is given
by the creation  rate
 defined in eqs. (\ref{Ndef})-(\ref{Ndeux}). The resulting equation for the
reduced density matrix of a quasi-particles with positive group velocity
(incoming) and chirality $\X$, is then given by:

\bea \len{\partial_t \rho(p,p';t)+i[H_{herm}, \rho(p,p';t)]=-2\g
\rho(p,p';t)+\frac 1 {\Xi L}\int dz_0 \theta(-z_0)\nn}\\ &
 \int_{-\infty}^{+\infty}dp_0  N(z_0, t, p_0)
 \widetilde\Psi_{inc}(p,t;t,p_0,z_0,d,\chi)
\widetilde\Psi^\dagger_{inc}(p',t;t,p_0,z_0,d,\chi).\label{vnc}\eea
 $\Xi$ has been defined in eq. (\ref{Ndef}), $L$ is the volume of the one
dimensional box in which the density matrix is normalised, and the
$\theta(-z_0)$ factor in the inhomogeneous term accounts for
 quasiparticle creation in the unbroken phase. $\Psi_{inc}$ was defined in eqs.
(\ref{onde})-(\ref{wpf}).
 The solution of eq. (\ref{vnc}) is

\bea\len{\rho(p,p';t)= \frac {2\g}{\Xi L}\int dz_0 dt_0 dp_0
\theta(-z_0)\theta(t-t_0)n_F{(\omega^0+\chi p_0/3,T)}\nn}\\  & & \qquad
 \widetilde\Psi_{inc}(p,t;t_0,p_0,z_0,d,\chi)
\widetilde\Psi^\dagger_{inc}(p',t;t_0,p_0,z_0,d,\chi),\label{rho}\eea
where eq. (\ref{Ndeux}) has been used.  From eq. (\ref{xwall}) in appendix
\ref{app-proof}  and eq. (\ref{densit})  it follows that

\be  \int dp \mbox{Tr}[\rho^u_\X(p,p)]=1. \ee

 It is possible to see that eq. (\ref{rho}) reduces to the equilibrium
distribution in the absence of the wall, i.e., when the factor $\theta(-z_0)$
is taken off.
Integrating then $z_0$ from $-\infty$ to $+\infty$ yields a $2 \pi\delta(p-p')$
factor, consistent with translation invariance in the absence of a wall,  and
$\rho$  becomes diagonal in momentum as expected. The $t_0$-integral cancels
the $2 \gamma$ factor in eq. (\ref{rho}),
\be\rho(p,p';t)=\frac{2\pi\delta(p-p')}{\Xi L}\int dp_0 n_F(\omega_\X+p_0/3,T)
  \frac{e^{-d^2(p-p_0)^2}}{\pi^{1/2}/d}.\label{roppp}
\ee

In the infinite volume limit, $2 \pi \delta(0)\simeq L$, and the $p_0$ integral
gives a gaussian smearing of the equilibrium distribution,
\be \rho(p,p;t)=\frac 1 \Xi \int dp_0 n_F(\omega_\X+p_0/3,T)
  \frac{e^{-d^2(p-p_0)^2}}{\pi^{1/2}/d}\,\,\, {\simeq\atop dT\gg 1}\,\,\,\frac
{n_F(\omega,T)}\Xi\label{ropp}\ee
 since the smearing can be neglected for $d$ large enough, i.e.
\be
T\gg\Delta E=\frac{dE}{dp_0}\Delta p_0\sim \frac{1}{3d},  \label{dcond}
\ee
which is the usual condition (\ref{condi}), (\ref{cond}), and the Fermi
equilibrium distribution is obtained.

In the present problem, the wall at $z=0$ requires  to treat separately the
$z_0<0$ and $z_0>0$ regions, though. Only the first one is involved in
describing the incoming flux, and the $z_0$ integration in eq. (\ref{rho})
provides an extra non-diagonal piece arising from the principal part of
$1/(p-p')$. This term is exponentially suppressed when
$|p-p'| > 1/d$, as (\ref{wpf}) imposes that both $p$ and $p'$  lie within a
distance $1/d$ from some $p_0$. $1/d$ is then an upper bound for the
non-diagonality of $\rho$. Depending on the shape and height of the wall, the
$z_0>0$ contribution may ultimately reduce the maximum $|p-p'|$, as it was
shown above when the wall was removed. This non diagonality of $\rho$ is a
quantum effect \footnote{A diagonal density matrix is just a probability
distribution, i.e. it is purely classical, it exhibits no quantum coherence.}
as is manifest in the oscillation of the $\exp{(-i (p-p')(t-t_0)/3)}$ factor
present in (\ref{rho}), through (\ref{wpf}) and (\ref{defpofw}).

A  simpler way to describe the effect follows by computing the Fourier
transform  of $\rho$ with respect to $(p-p')$, in order to obtain the Wigner
function $f(p,z)$ \cite{balescu}. A sharp cutoff imposed in $z_0$ then
translates into inhomogeneities of $f(z,p)$ on scales at least bigger than $d$,
the individual particle diameter.
The fact that $f(z,p)$ is not a real positive function (as it would be in the
semiclassical limit) is a sign of quantum effects.

\subsubsection{Reflection probability from the density matrix.}

The thermal average of any physical observable $O$ is given by $\mbox{Tr}(\rho
O)/\mbox{Tr}(\rho)$. In particular, the reflection probability is  the thermal
average of $r_\X(p_\X)r^\ast_\X(p'_\X)$, with $r_\X(p_\X)$ as defined in
(\ref{refcoeff}). For a diagonal density matrix, i.e. in the absence of quantum
coherence, the reflection probability amounts to the average of $\vert
r_\X(p_\X)\vert^2$, and this is what happens
for vanishing damping rate
(see the discussion in subsection (\ref {sec-refprob})).

The density of incoming particles at position $z<0$ is given in terms of the
density matrix by
\be d^{inc}(z)=\frac 1 {2\pi} \int\int dp dp' e^{i (p-p')z}\rho(p,p')=\frac 1
L\quad\mbox{if}\quad 0<-\frac z d \ll 1.\label{density}\ee
In fact $d^{inc}(z)$ depends only slightly on $z$ when $3\g d\ll1$:
\be d^{inc}(0)= \frac {\Xi^{wall}}{L \Xi}\simeq \frac 1 L \quad\mbox{if}\quad
3\g d \ll 1,\ee
as can be seen from eqs (\ref{xwall}) and (\ref{belle}) in appendix
\ref{app-proof}.
 For the problem under study,  with a non diagonal density matrix, the
reflection probability is given by the ratio of the density of reflected
particles over the incoming density, i.e.

\be n_r=\frac L {2\pi}\int dp dp' \rho(p,p';t=0) r(p) r^\dagger(p'),
\label{trhor}\ee
where  time translational invariance  was used to choose $t=0$. As expected,
eq. (\ref{trhor}) is equivalent to (\ref{refprob}) and the calculation will
proceed exactly as in the preceding
subsections.

\seCtion{Several flavours: CP violation.}
\label{sec-CP}
In this section, we include the non-diagonal weak contributions to the
effective Hamiltonian in
a perturbative expansion. We then compute the CP-asymmetry and compare our
results with
existing literature.

\subsection{Effective Hamiltonian}
\label{sec-specsf}

Only the quarks with the same charge are mixed,
and the Hamiltonian will be factorised for down and up sectors.
In section (\ref{sec-spec}), we found the effective unperturbed Hamiltonian
(\ref{heffo2}) in the case of several generations, which we recall here:

\be
H^0_{eff}=\left(\begin{array}{cc
} -\frac{1}{3} i \sigma_z\partial_z   -i \gamma  + {\hat{\omega}}^0_R &
\frac{m}{2} \theta(z) \\ \frac{m}{2} \theta(z) & \frac{1}{3} i
\sigma_z\partial_z  -i \gamma +{\hat{\omega}}^0_L \end{array}
\right).
\ee

It contains only the leading $T$, QCD and flavour diagonal electroweak
corrections. It has been linearised in momentum.
In order to obtain an observable asymmetry, the remaining corrections,
i.e. the non-diagonal weak corrections and those subleading in T, are essential
ingredients. The full effective hamiltonian is

\be H_{eff}= H^0_{eff} + \theta(-z) \delta H^u_{eff} + \theta(z) \delta
H^b_{eff},
\label{htot}\ee
where
\be
\delta H^{u,b}_{eff} =  \frac{1}{2}\left(\begin{array}{cc
}  [\delta h^{u,b}_R(\omega,k) -i\delta a_R^{u,b}(\omega,k) \sigma_z\partial_z
]&\frac{1}{2} m c(\omega,k) \theta(z) \\  \frac{1}{2} m c(\omega,k) \theta(z) &
 [\delta h^{u,b}_L(\omega,k) +i\delta a^{u,b}_L(\omega,k) \sigma_z\partial_z
]\end{array}
\right).
\label{hefftot}
\ee
$\delta h^{u,b} = h^{u,b} - \bar{h}$ and $\delta a^{u,b} = a^{u,b} -\bar{a}$
are flavour dependent and non-diagonal, as defined and discussed in subsections
(\ref{sec-realun}), (\ref{sec-asun}) and (\ref{sec-several}). They have to be
also linearised in momentum.
To be consistent with our approximation (i.e. that these effects do not
modify the residues derived from the free hamiltonian (\ref{heffo}))
we must expand these functions around the poles (\ref{poles}) and take
only the zero order. This means that  corrections of $O( k \delta h,a,c)$ are
neglected, which is a reasonable assumption for the low momentum region we are
considering.

Finally, the hamiltonian we must consider is:

\be
H_{eff}=H^0_{eff} + \frac{1}{2}\theta(-z) \left(\begin{array}{cc
} \delta h^u_R(\omega^0,0) & 0 \\ 0  &  \delta h^{u}_L(\omega,0) \end{array}
\right)+ \frac{1}{2}\theta(z) \left(\begin{array}{cc
} \delta h^b_R(\omega^0,0) & m  c({\omega}^0,0) \\ m c({\omega}^0,0)  &  \delta
h^{b}_L(\omega^0,0) \end{array}
\right),
\label{heff}
\ee
where we can take $\omega^0 = \omega^0_{QCD}$, i.e., it contains just the
dominant QCD part.
The precise choice of $\omega^0$ in the arguments of the perturbed functions
only influences subdominant contributions to the asymmetry.

\subsection{Reflection Matrix and CP-Asymmetry}
\label{sec-resf}

We are now ready to compute the CP-asymmetry defined in Sec. 2, eq.
(\ref{dcp}),
\bea
\Delta_{CP} = \frac L {2\pi}\mbox{sign}(\chi) \int \int dp\,dp'\,[\; \rho \;
\mbox{Tr}\{{r^u_{-\chi}}^\dagger r^u_{-\chi} -
{({\bar{r}}^u_{\chi})}^\dagger{{\bar{r}}^u_{\chi}}\} ].
\eea
Being a CP-odd asymmetry it must contain the contribution of CP-violating
phases, for which at least three generations are needed. In this case,
the reflection and transmission amplitudes $r, t$ are $3\times 3$ matrices in
flavour space.

We show below that an effect is present al order $\alpha_W^2$ in rate, through
subleading $T$ corrections. At leading order in $T$, $ O (T^2)$, there is no
contribution at this electroweak order, because the only flavour dependence is
through Yukawa couplings. The physical reason is CKM unitarity. The order
$\alpha_W^2$ effect is tantamount to a one-loop electroweak correction in the
amplitudes. Denote by $i$ and $f$ the external initial and final flavours, and
$l$ (or $l'$) the internal one in the loop. Convoluting two of these diagrams
gives a contribution to the rate $ \propto\,Im \{ K^*_{lf} K_{li}(K^*_{l'f}
K_{l'i})^*\}$.If the function accompanying this factor contains no
antisymmetric dependence on the internal flavours, the contribution will
vanish when the sum over $l$ and $l'$ is performed. At leading order in $T$,
the dependence on Yukawa couplings is insufficient to produce a non-zero
effect, because they factorise as a symmetric function of
$l,l'$. $\Delta_{CP}$ would then be a higher order electroweak effect, i.e.,
$\alpha_W^3$. Subleading $T$ corrections do contain an explicit internal quark
mass dependence, which results in the above mentioned antisymmetric functions.

In section 4, we have shown that the reflection probability of quasi-particles
in a single flavour world can be written  as a gaussian
smear-out of the reflection probability for planes waves (with zero damping
rate) after making an analytic
continuation $p_\X \ra p_\X + 3 i \g$, eq. (\ref{nrsimpl}). This result holds
as well in the case of
several generations:
\bea \len {n^\X _r(0,0)= \frac 1{\Xi}\int_{-\infty}^{+\infty} dp_0
n_F{(\omega_\chi^0 + p_0/3, T)}\nn} \\ & &
\frac d {\sqrt{\pi}}\int_{-\infty}^{\infty} dp_\X e^{-d^2(p_\X-p_0)^2
+(3d\g)^2} \mbox{Tr}[r_\chi(\pi_\X) r_{\chi}^\dagger(\pi_\X)],
\label{R2sev}
\eea
where now $r_\chi(\pi_\X)$ are $3\times 3$ matrices and Tr refers to
the tracing over the flavour indices. The problem is then reduced to calculate
the reflection matrices for plane waves $r_\chi(p_\chi) = r_\chi(\omega)$,
evaluate them in the complex point $\pi_\X$ and, finally, perform the gaussian
and thermal averages.

Let us start by computing $r_\chi(\omega)$.
 For $\delta h = 0$, the problem reduces to three independent one-flavour
cases,
whose solution we know, eq. (\ref{refcoeff}).
An analytic expression in the case of the simplified effective Hamiltonian, eq.
(\ref{heff}),
is still too difficult to find. We will use a
perturbative approach to obtain an analytic result. The procedure
involves deriving an exact equation for the reflection amplitude, and
proceeding then by iteration,
in powers of $\delta h_{L,R}$. This corresponds to an expansion of the
asymmetry (\ref{dcp}) in
$\alpha_w sin \theta_c$, as the CP-violating effects are contained in the
non-diagonal weak terms.

We work in the flavour basis that diagonalises the unbroken effective
Hamiltonian, $H_{eff}^u$.  The transformation
to this basis from the one corresponding to the unperturbed Hamiltonian,
$H^0_{eff}$, has the form

\be
\psi \equiv \left(\begin{array}{c
} \psi^R \\ \psi^L \end{array}
\right) =
\left(\begin{array}{cc
} I_3 & 0 \\ 0  &  O_L \end{array}
\right) \left(\begin{array}{c
} \psi'^R \\ \psi'^L \end{array}
\right) \equiv O \psi',
\ee
where $\psi^\chi$ are spinors with three flavour components, $I_3$ is the
identity matrix and $O_L$ is
a unitary $3 \times 3$ matrix. In the prime basis, the solution for right
quasiparticle with energy $\omega$ coming from the unbroken phase has the usual
form (\ref{planew}):
\bea
\psi'^\chi_{inc}(z,t;\omega)=e^{-i \omega t} & \bigg\{ &
  \theta(-z)\left[e^{ i p_\chi    z}                u_{ \chi,\chi}+
                  e^{-i p_{-\chi}z} r_\chi(p_\chi) u_{-\chi,\chi}
            \right]\nn\\
 & + &\theta(z) e^{i p^t_\chi z}
   [u_{\chi,\chi} + r_\X(p_\chi) u_{-\chi,\chi}] \; \bigg\},
\eea
where $p_\chi$ is the incoming particle momentum as defined in eq.
(\ref{pichi}).

It is possible to obtain an implicit equation for $r_{R}$ and $r_{L}$ by
imposing,

\be
z>0   \;\;\;\;\;\;\;\;\; O^\dagger H^b_{eff} O \;{\psi}' = \omega \;{\psi}'.
\label{dfs}
\ee

For an incoming right-handed quark, eq. (\ref{dfs}) gives
\bea
r'_{R} [{{\hat{\omega}}_R}^0 -\omega + \frac{1}{2} \delta h^b_R(\omega^0)]
+ [{{\hat{\omega}}_L}^0 -\omega + \frac{1}{2} \delta h^b_L(\omega^0)] r'_{R}
\nonumber
\eea
\be
+ \frac{1}{2} r'_{R} [\;1-c(\omega^0)\;] m r'_{R}
+\frac{1}{2} m [\;1-c(\omega^0)\;]^\dagger = 0,
\label{FS}
\ee
where ${r'}_{R} \equiv O_L r_{R}$. As $O_L$ is unitary, $\Delta_{CP}$ is
unchanged upon replacing $r \rightarrow r'$.

In the same way, the equation for antiparticles reads
\bea
{\bar{r}}'_{R} [{{\hat{\omega}}_R}^0 -\omega + \frac{1}{2} (\delta
h^b_R(\omega^0))^*]
+ [{{\hat{\omega}}_L}^0 - \omega + \frac{1}{2} (\delta h^b_L(\omega^0))^*]
{\bar{r}}'_{R}
\nonumber
\eea
\be
+ \frac{1}{2} {\bar{r}}'_{R} [\;1-c(\omega^0)^*\;] m {\bar{r}}'_{R}
+\frac{1}{2} m [\;1-c(\omega^0)^*\;]^\dagger = 0.
\label{FSa}
\ee
A similar equation for $L$ reflection amplitudes can be written, although it is
not needed to compute $\Delta_{CP}$.

We
look for a perturbative solution of equations (\ref{FS}) and (\ref{FSa})
in powers of $\alpha_w$:
\be
r_{R} = r^{(0)}_{R} + r^{(1)}_{R} + r^{(2)}_{R} +...
\label{rseries}
\ee
$r^0_{R}$ is then the solution in the limit $\delta h,\delta a, c =0$ and
it reduces to (\ref{refcoeff}), as expected. To lighten the notation,
we will skip the subindices $R$ on $r$'s from now on. The first and second
orders are given by
\bea
r^{(1)}_{ij} = - \frac{1}{2} \frac{(\,r^{(0)}{\delta h^b}^{(1)}_R + {\delta
h^b}^{(1)}_L r^{(0)} - r^{(0)} {\delta c}^{(1)} m  r^{(0)}
- m \delta c \,)_{ij}}{d_{ij}}\nonumber\\
\eea
and
\bea
r_{ij}^{(2)} = - \frac{1}{2} \frac{(\,r^{(1)}{\delta h^b}^{(1)}_R + {\delta
h^b}^{(1)}_L r^{(1)} - r^{(1)} {\delta c}^{(1)} m  r^{(0)}
- r^{(0)} {\delta c}^{(1)} m  r^{(1)}\,)_{ij}}{d_{ij}} \nonumber\\
-\frac{1}{2} \frac{(\, r^{(1)} m r^{(1)} + r^{(0)}{\delta h^b}^{(2)}_R +
{\delta h^b}^{(2)}_L r^{(0)} - r^{(0)} {\delta c}^{(2)} m  r^{(0)}
- m {\delta c}^{(2)}\,)_{ij}}{d_{ij}} ,
\eea
where the indices refer to flavour and $d_{ij} \equiv (\omega^i_L+\omega^j_R -
2 \omega) +
\frac{m_i r^0_{ii}}{2} + \frac{m_j r^0_{jj}}{2}$.

It follows from the discussion
in subsection 4.1 that $r^0$ are complex in the region of total reflection, and
this region is the same for particles and antiparticles. This gives rise to
CP-even phases. In order to generate an observable CP-asymmetry,
they must interfere
with the CKM one, contained in the non-diagonal weak corrections: $\delta
h_R^b$, $\delta h_L^b$, $c$.
Using eq. (\ref{hh}), and neglecting those terms in $\delta h_L$ that
have the same dependence on internal masses (or Yukawa couplings) than $\delta
h_R$, because they
give a zero contribution at this order
, we can easily obtain:
\bea
\delta h_R^b = \alpha_w \lambda_i \lambda_f \sum_{l} K_{li} K^*_{lf}
I_R(M_l^2),
\;\;\;\;  \delta h_L^b = \alpha_w \sum_{l} K_{li} K^*_{lf} I_L(M_l^2)
\eea
and
\bea
c = \frac{\lambda_f}{m_i} \sum_{l} K_{li} K^*_{lf} I_m(M_l^2),
\eea
where we have defined
\be
I_R(M_l^2) = - \frac{\pi}{2} H(M_l,M_W), \;\; I_L(M_l^2) = \lambda_l^2
I_R(M_l^2), \;\;
I_m(M_l^2) = \pi \lambda_l M_l C(M_l,M_W).
\ee

It then follows that the first effect in
the asymmetry appears at $O(\alpha_w^2)$ and it comes only
from the interference of the $O(\alpha_w)$ effects in $\delta h_R^b$
and $\delta h_L^b$. Consequently, there is no effect at $O(\alpha_w^2)$
at leading order in $T$, because at this order $\delta h_R^b = 0$.
It is interesting to analyze the expression for the non-integrated
asymmetry at this
order, where the GIM mechanism is explicitly operative:

\bea 
\Delta^{(2)}_{CP} \equiv Tr[\;{r^{(1)}}^\dagger r^{(1)} + {r^{(2)}}^\dagger r^{(0)} 
+ {r^{(0)}}^\dagger r^{(2)} - antiparticles\;]\nonumber\\
\sim \sum_{i,j} Im[\;\delta h_L^b )_{ji} \delta h_R^b )_{ij}] \times
Im\{{r^0_{ii}}^* [ \,\frac{r^0_{jj}}{|d_{ij}|^2}+  \frac{m_j ((r^0_{ii})^2 
-(r^0_{jj})^2 )}{2 d_{ii} d_{ij} d_{ji}}  + \frac{ r^0_{jj}}{d_{ii}} (\frac{1}{d_{ij}} + \frac{1}{d_{ji}})\, ]\;\}.\nonumber\\
\label{series}
\eea
$\Delta^{(2)}_{CP}$ can be shown to have the following structure:
\be
\Delta^{(2)}_{CP} \sim \alpha_w^2 \;\;(2 i J)\;\; T^{int}\;\; T^{ext},
\label{jardelta}
\ee
where $J$, $T^{int}$ and $T^{ext}$ contain the expected ``\`a la Jarlskog''
behaviour of the asymmetry as a function of the weak angles
($J$), the internal quark ($T^{int}$) and the external quark masses
($T^{ext}$). The connection between (\ref{series}) and (\ref{jardelta})
is

$$
Im[\delta h_L^b )_{ji} \delta h_R^b )_{ij}] = \alpha_w^2 \lambda_i \lambda_j
2 i \sum_{l,l'} Im[K_{li} K^*_{lj} K_{l'j} K^*_{l'i}] (\lambda_l^2
- \lambda_{l'}^2) I_R(M_{l'}^2) I_R(M_l^2)$$\be
\equiv \alpha_w^2 \lambda_i \lambda_j (\pm 2 i J) T^{int} ,
\ee
with
\bea
J \equiv   \pm Im[K_{li} K^*_{lj} K_{l'j} K^*_{l'i}] = c_1 c_2 c_3 s_1^2 s_2
s_3 s_\delta,
\nonumber
\eea
and
\bea
T^{int} \equiv \sum_{l} (\lambda_l^2
- \lambda_{l+1}^2) I_R(M_l^2) I_R(M_{l+1}).
\eea
The $\pm$ sign in $J$ refers to the cyclic order of $i,j$
and $l+1$ must be understood as modulo 3. $T^{int}$ is
the oriented area of the triangle formed by the three points in the complex
plane given by $\lambda_l^2 I_R(M_l^2) + i I_R(M_l^2)$, corresponding to the
three flavours of the internal quarks: $l=1,2,3$, shown in fig. (\ref{fig-tri}
(a))
{}From this construction, it is obvious that the triangle would collapse into
a line whenever two internal masses are degenerate.

Finally,
\be
T^{ext} \equiv \sum_{i,j} \lambda_i \lambda_j
Im\{{r^0_{ii}}^* [ \,\frac{r^0_{jj}}{|d_{ij}|^2}+  \frac{m_j ((r^0_{ii})^2
-(r^0_{jj})^2 )}{2 d_{ii} d_{ij} d_{ji}}  + \frac{ r^0_{jj}}{d_{ii}}
(\frac{1}{d_{ij}} + \frac{1}{d_{ji}})\, ]\;\}.
\ee

Although this expression is more involved, it vanishes whenever two external
quarks masses are degenerate. This can be seen explicitly in a triangle
construction for the
analytic continuation to $\omega + i \gamma$, eq. (\ref{nrsimpl}), in the limit
$m<< \gamma$. Then,
\be
d_{ij} \sim d_0 + O(\frac{m}{\gamma}) = - 2 (i \gamma + \omega) + \omega_R^o +
\omega_L^o,
\ee
leading to
\be
T^{ext} \sim \frac{|d_0|^2+2 Re(d_0^2)}{|d_0|^4} \sum_i \lambda_i \lambda_{i+1}
(r^0_{i+1 i+1} {r^0_{ii}}^* - r^0_{ii} {r^0}^*_{i+1 i+1}),
\ee
where again the sum over $i$ is modulo 3.
$T^{ext}$ is proportional to the oriented area
of a triangle with vertices in the complex plane given by $\lambda_i r^0_{ii}$,
Fig. \ref{fig-tri}-(b). When any two external quarks masses are degenerate, two
of
these points would coincide and the area would vanish. This shows that
the GIM cancellation also works explicitly for external quark masses, as
expected.

\begin{figure}[t]   
    \begin{center} \setlength{\unitlength}{1truecm} \begin{picture}(5.0,3.0)
\put(-6.0,-4.0){\includegraphics{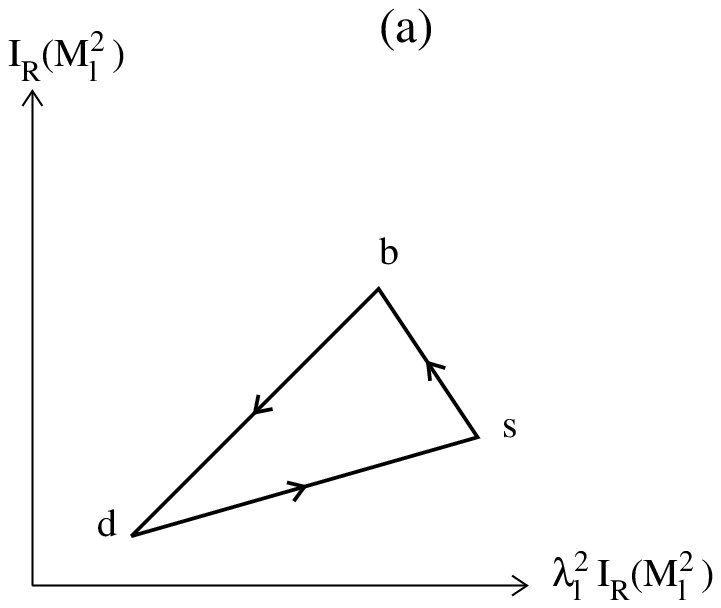}}
\put(3.0,-4.0){\includegraphics{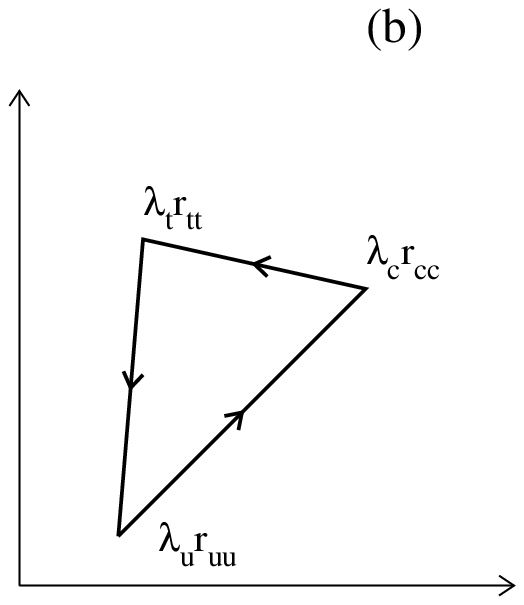}}

       \end{picture} \end{center} 	\vskip 2.6cm \caption[]{\it{ (a) The
triangle in the complex plane, with area $T^{int}$. (b) The triangle in the
complex plane, with area $T^{ext}$. In both cases, the sign is negative
(positive) if the arrows turn (anti)-clockwise.
This figure is only illustrative,  a realistic triangle would look extremely
flat due to the huge mass hierarchy between quark masses.
 }} \protect\label{fig-tri}
\end{figure}
\subsection{Results and Comparison with literature}
\label{results}

We now give the numerical results we have obtained at $O(\alpha_w^2)$
in the reflected baryonic flux, both for up external quarks (u,c,t) and downs
(d,s,b).
We have used the following values for the masses in GeV, $M_W=50$, $M_Z=57$,
$m_d=0.006$, $m_s=0.09$, $m_b=3.1$, $m_u=0.003$, $m_c=1.0$, $m_t=93.7$. The
couplings are $\lambda_d=1.2\,10^{-4}$, $\lambda_s=1.8\,10^{-3}$,
$\lambda_b=6.2\,10^{-2}$, $\lambda_u=6.2\,10^{-5}$,
$\lambda_c=2\,10^{-2}$ and  $\lambda_t=1.88$, and $\alpha_s=0.1$,
$\alpha_W=0.035$. Fig. (\ref{alphatwo}) shows $\Delta(\omega)$ for the ups (the
dominant contribution). The result for the averaged ensemble asymmetry is,
\be
{\Delta^{uct}_{CP}}= 1.6\,\, 10^{-21},\qquad\qquad
{\Delta_{CP}^{dbs}}= -3\,\,10^{-24}. \label{resultu}\ee
In both cases, we find that the asymmetry is dominated by the two heavier
external quarks.

\begin{figure}
\hbox to \hsize{\hss\psboxto(0.48 \hsize;0pt){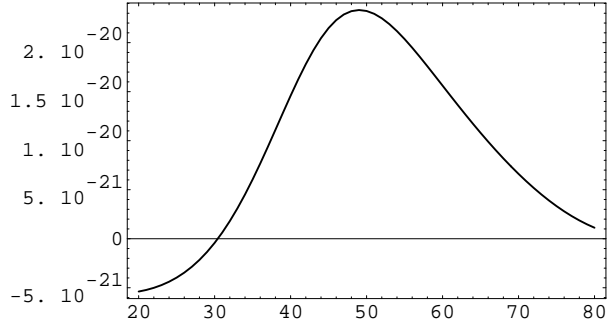}\hss}
\caption{
The dominant ($O(\alpha^2_W)$) non-integrated CP asymmetry,
$\Delta_{CP}(\omega)$, when mass effects are included
inside thermal loops, as a function of the quasi-particle energy, $\omega$. The
figure corresponds to charge 2/3
flavours.}
\protect\label{alphatwo}
\end{figure}
These results are several orders of magnitude in disagreement with
the previous results obtained by Farrar and Shaposnikov in ref. \cite{shapo}.
Their estimate is $\Delta_{CP} \,\stackrel{>}{\sim}\,10\,^{-8}$ (see their eq.
(10.3)). This tremendous discrepancy can fortunately be
understood. There are mainly two important effects these authors did not
consider. The first one, which explains the orders of magnitude
difference, is that they disregard the effect of the damping rate in the
scattering of the particles on the wall. In our calculation, however, it is an
essential ingredient.
The second difference, responsible for a small factor in the
discrepancy, is due to the fact that they just considered the
leading-T corrections in the self-energies.
We have seen in the previous section that
 the subleading terms  contain the dependence on the internal masses, essential
for CP-violation. When they are neglected
no $O(\alpha_w^2)$ effect can appear, and
the first contribution would loom at $O(\alpha_w^3)$.

In order to prove that these latter considerations are correct, we have
reproduced
their results, i.e. we have computed the CP-asymmetry just considering
leading-T  corrections in (\ref{heff}) and setting $\gamma = 0$. The result
is shown in Fig. (\ref{alphathree}-a) and it is in complete agreement with the
results in \cite{shapo}
(see Fig.5 in this reference). Improving their numerical calculation by the
inclusion of the damping rate, gives the result shown in Fig.
(\ref{alphathree}-b). The integrated asymmetry from fig (\ref{alphathree}-b)
 is $\sim 4\cdot 10^{-22}$, to be compared to
the result in fig  (\ref{alphathree}-a) of \cite{shapo}, $\sim 10^{-8}$.

\begin{figure}
\hbox to \hsize{
\vbox{\psboxto(0.48\hsize;0pt){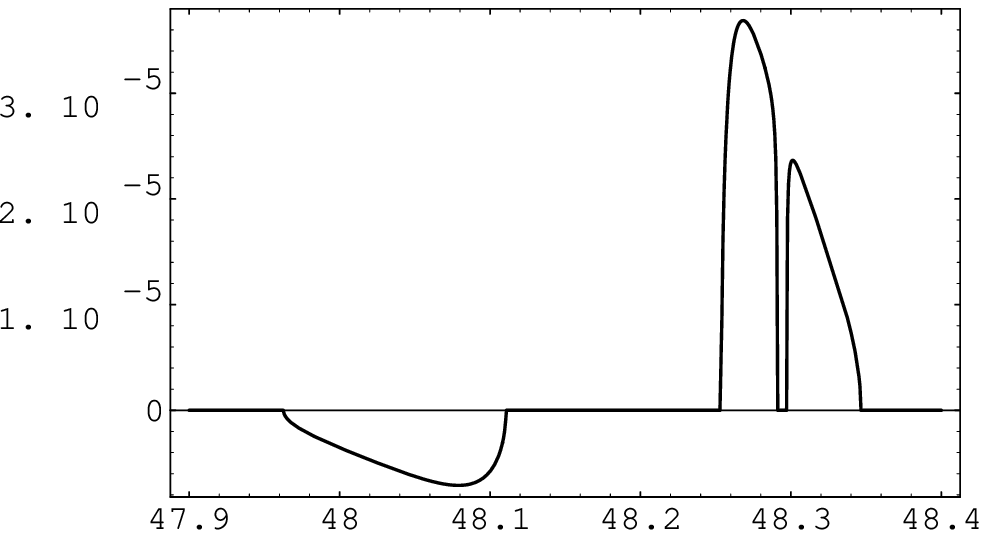}
      \hbox to \drawingwd{\hss(a)\hss}}%
      \hss
\vbox{\psboxto(0.48\hsize;0pt){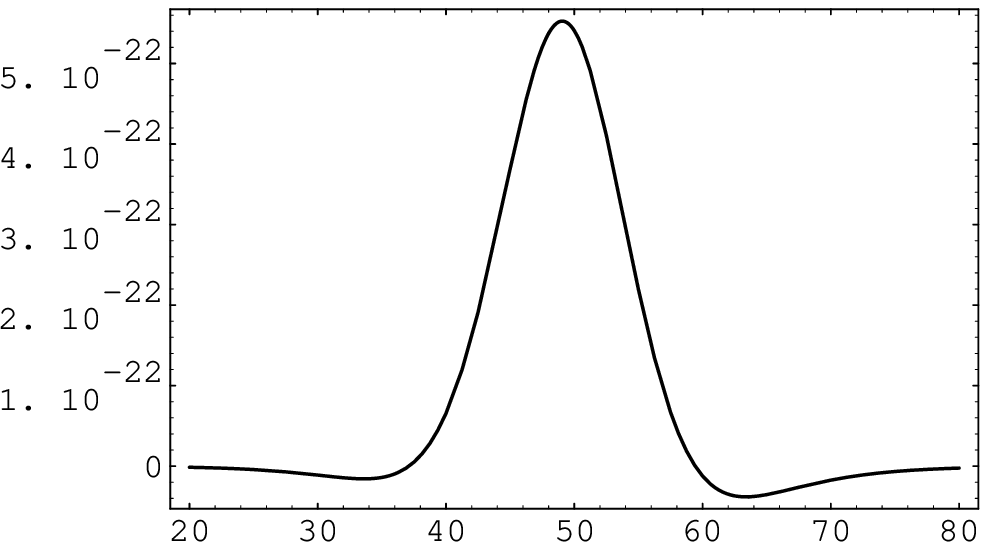}
      \hbox to \drawingwd{\hss(b)\hss}}}
\caption{(a) shows the non-integrated CP asymmetry ($\Delta_{CP}$) produced by
down quarks in the narrow energy range which dominates for zero
damping rate, when masses are neglected in the internal loop.  (b)
shows the dramatic effect of turning on the damping rate effects, in
the same approximation.}
\protect\label{alphathree}
\end{figure}

 With or
without the damping rate, it is also possible to work out an analytic
expression  at third order, from the perturbative
calculation (\ref{rseries}),(\ref{series}). For $\gamma = 0$,
the expected GIM cancellation appears, although there
is no simple formula that explicitly shows them.
On the other hand, in the case $\gamma \ne 0$ and in the
limit $m << \gamma$ \footnote{This is valid for down external quarks, the case
we
considered}, the expression for the peak value of the asymmetry
beautifully reduces to

\be
\Delta_{CP}^{max} =
\left[\sqrt{\frac{3\pi}{2}} \frac {\alpha_W
T}{32\sqrt{\alpha_s}}\right]^3
J\,\frac{(m_t^2-m_c^2)(m_t^2-m_u^2)(m_c^2-m_u^2)}
{M_W^6}\,
\frac{(m_b^2-m_s^2)(m_s^2-m_d^2)(m_b^2-m_d^2)}{(2\gamma)^9}\nn\\
\label{D3}
\ee
This was expected from naive order-of-magnitude arguments.

Finally, the  results  (\ref{resultu}) show that non-leading effects
in T give the main contribution to the asymmetry in the case of non-vanishing
damping rate and, in contrast with \cite{shapo},  the up-sector dominates the
asymmetry.

Very recently, Huet and Sather\cite{slac} have analyzed the problem. These
authors state that
they confirm our conclusions. As we had done in ref. \cite{letter}, they stress
that the damping rate is a source for quantum decoherence, and use as well an
effective Dirac equation which takes it into account. They discuss a nice
physical analogy with the microscopic theory of reflection of light.
They do not use wave packets to solve the scattering
problem, but spatially damped waves, as in our heuristic treatment at the
beginning of Sect. \ref{sec-wall}.

\subsection{Wall thickness.}
\label{sec-thick}

Notice that the derivation in sect. \ref{sec-wall} is totally independent of
the shape of the function $r(k)$. The only requirement was a singularity
structure limited to a cut in the region of total reflection. This is quite
generic: only for very special wall shapes can other singularities be
expected. For instance, when the wall is not monotonous, a pole with an
imaginary part may express the decay of a quasi-bound state trapped in a
potential well.

 The thin wall approximation used in this paper is valid only for wall
thickness $l\ll1/6\gamma$, while perturbative estimates suggest
$l\ge.1$GeV$^{-1} \ge 1/6\gamma$. The CP asymmetry, generated in the presence
of such more realistic walls, would be orders of magnitude below the thin wall
estimate, eq.  (\ref{resultu}), reinforcing thus our conclusions, because a
quasi-particle would then collide and loose coherence long before feeling a
wall effect.

\seCtion{Conclusions.}
 \label{sec-conclu}

At finite temperature a plasma is mainly an incoherent mixture of states. The
scattering of quasi-quarks with thermal gluons induces a large damping rate,
$\gamma$. Quantum coherence is only maintained between two inelastic
scattering processes of this type. It is lost over spatial distances larger
than $\sim 1/6\gamma\,\sim (120$ GeV$)^{-1}$, and over time scales larger than
$\sim1/2\gamma\,\sim\, (40$ GeV$)^{-1}$.

We have considered both an heuristic approach to quasi-quarks in terms of
spatially damped waves, and a more physically sound one in terms of wave
packets which incorporates the above mentioned characteristics of coherent
mean free time and coherence length. With these tools, we have shown that
coherent tree-level reflection of quasi-quarks hitting the bubble wall is
suppressed for all flavours but the top, by a factor $m/2\gamma$. This
effect might also be relevant in certain non-relativistic systems, e.g. in
solid state physics. We have shown as well that a CP-violation
reflection asymmetry appears already at order $\alpha_W$ in amplitude, and it
is suppressed by further powers of $m/2\gamma$. It fails to explain the baryon
asymmetry by more than 12 orders of magnitude.

Any non-standard electroweak scenario for baryogenesis, where quantum coherence
is required over distances and times larger than the above mentioned ones, is
subject to the same flaw. In particular, ``realistic'' electroweak walls are
thick, much larger than the coherence length. No (quark) coherence should be
required through the whole wall thickness in a successful scenario. In a more
positive view, it is worth to stress that promising candidate theories are
those where CP violation is associated to the top and/or heavier flavours.
Leptonic induced baryogenesis may also be safe.

\seCtion{Acknowledgements.}

We acknowledge  Tanguy Altherr, Luis Alvarez-Gaum\'e, Philippe Boucaud,
 Andy Cohen, Alvaro De R\'ujula, Savas Dimopoulos,  Jean Marie Fr\`ere, Jean
Ginibre, Gian Giudice, Jean-Pierre Leroy, Manolo Lozano, Jean-Yves Ollitreault,
Anton Rebhan,  Eric Sather, Dominique Schiff  and Javier
Vegas for many inspiring
discussions.  We are indebted to Nuria Rius for criticisms on the typescript.
M.B. Gavela and P. Hernandez are indebted to ITP (Santa Barbara) for
hospitality during the final period of this work, where their research was
supported in part by the National Science Foundation under Grant No.
PHY89-04035. C. Quimbay would like to thank COLCIENCIAS (Colombia) for
financial support. P.H. acknowledges partial financial support from
NSF-PHY92-18167 and the Milton Fund.

\appendix

\seCtion{Proof of the simplified formula for the reflection probability.}
\label{app-proof}

This appendix develops a  bound to the error  introduced in eq. (\ref{nrcomp})
when
the integration $\int_{-\infty}^0 dt_0$ in eq. (\ref{refprob}) is replaced by
$\int_{-\infty}^{+\infty} dt_0$. We  derive an upper bound
of $u^\dagger{-\X,\X}$ $\Psi_{ref}(0,0;z_0,t_0,p_0,d,\chi)$\footnote{The factor
$u^\dagger{-\X,\X}$  takes care of the spinor $u{-\X,\X}$ in $\Psi_{ref}$, so
as to keep only the spatial dependence.}. A comparison of eqs. (\ref{onde}) and
(\ref{wpr}) shows that:

\bea \len{
u^\dagger_{-\X,\X}\Psi_{ref}(z,t;z_0,t_0,p_0,d,\chi)=\nn}\\ & &\frac 1 {\sqrt{2
\pi}}
u^\dagger_{\X,\X}\int_{-\infty}^{+\infty} dp_\X e^{-i p_{-\X}z}r_\X(p_\X)
\widetilde \Psi_{inc}(p_\chi,t;z_0,t_0,p_0,d,\chi)\label{refinc}\eea

As $p_{-\X}=p_\X + 3(\omega^0_\X-\omega^0_{-\X})$, eq. (\ref{refinc}) can be
expressed as a  convolution,

$$
u^\dagger_{-\X,\X}\Psi_{ref}(z,t;z_0,t_0,p_0,d,\chi)=$$\be
u^\dagger_{\X,\X}
\int dz_1
e^{-3i(\omega^0_\X-\omega^0_{-\X})z}\Psi_{inc}(z_1,t;z_0,t_0,p_0,d,\chi) \tilde
r_\X(z+z_1)\label{convol}
\ee
where
\be \tilde r_\X(z)= \frac 1 {\sqrt{2\pi}}\int e^{-ip_\X
z}r_\X(p_\X)\label{rtilde}\ee

The proof goes as follows. We first show that

\be \tilde r_\X(z) = - i \sqrt{\frac 2 \pi}\theta(z)\int_{-\pi/2}^{+\pi/2}
d\theta \cos(m z\sin \theta) \cos^2\theta\label{rtilde3}\ee
where $m$ is the quark mass. This gives the bound

\be \vert\tilde r_\X(z)\vert \le \sqrt{\frac \pi 2}m\theta(z)
.\label{rborne}\ee
 Next, this bound is inserted in eq. (\ref{convol}) and  the final bound
 on the $ t_0 >0$ contribution to the reflection probability in eq.
(\ref{refprob}) follows.

\begin{itemize}
\item Bound on $\tilde r_\X(z)$.

The integral in (\ref{rtilde}) is computed using (\ref{refcoeff}) and the
analytic properties of $r_\X(p_\X)$, explained after eq. (\ref{refprob}). For
$z<0$, the contour can be closed on the upper half plane,
with a vanishing result. Else, the cut has to be crosses with the result:

\bea \int dp_\X e^{-ip_\X z}r_\X(p_\X) &=& \theta(z) \int_{-m}^{+m} dp_\X
 e^{-ip_\X z} \left(\scriptstyle{ \frac m {p_\X +i \sqrt{m^2-p_\X^2}}-
\frac m {p_\X -i \sqrt{m^2-p_\X^2}}}\right)\\ \nn
	&=& -\frac{i m}2 \int_{-\pi/2}^{+\pi/2}
d\theta \cos(m z\sin \theta) \cos^2\theta\label{borne2}\eea
where a change of variables, $p_\X \ra m \sin\theta$, was performed. This
completes the proof of eq. (\ref{rtilde3}), and hence of eq. (\ref{rborne}).

\item Integrals on $z_0$ and $t_0>0$ for the incoming wave packet.

Before considering the effect of extending the integration on $t_0$ in eq.
(\ref{refprob}) to $+\infty$, we discuss the simpler exercise of integrating
the squared norm of the incoming wave $\Psi_{inc}$. The particle density $\Xi$
in  eq. (\ref{Ndef}) has been computed far from the wall in the unbroken
region. For that reason, the integral on $z_0$ extends to $+\infty$. Consider
the same density close to the wall, i.e. for $z=0^-$ (and $t=0$ to be
specific). Since only  packets created in the unbroken region are relevant, we
must now constrain the integral on $z_0$ from $-\infty$ to $0$. This  is the
density of incoming quasi-particles hitting the wall, $\Xi^{wall}$:

$$ \Xi^{wall}=\int dp_0 dt_0 dz_0 \theta(-z_0)\theta(-t_0)\vert
\Psi_{inc}^\chi(0,0;z_0,t_0,p_0,d)\vert^2 N(z_0,t_0,p_0) $$
\be<
\frac 1 {2\gamma} \int dp_0 N(z_0,t_0,p_0)= \Xi. \label{xwall}\ee
where the inequality stems from the positivity of the integrand and the
comparison with the integral in eq. (\ref{Ndef}). It is useful to notice that,
if the integral over $t_0$
 in the first line of eq. (\ref{xwall}) is extended
 to $+\infty$, we recover $\Xi$ i.e. the same result as extending the integral
over $z_0$ to $+\infty$. Let us estimate the difference between $\Xi^{wall}$
and $\Xi$, i.e. the effect of the integral over $t_0>0$ on the incoming
density. Afterwards we will extend the result to the reflected probability.

{}From the last line of eq. (\ref{wps}), it follows that

\bea \Xi- \Xi^{wall}=&\frac 1 {d\pi^{1/2}}\int_{-\infty}^{+\infty}dp_0
\int_{-\infty}^{+\infty} dx\int_{0}^{3x} dt_0 e^{-\frac {x^2}{d^2}+2\g
t_0}N(z_0, t_0,p_0)\\ \nn
=& \frac 1 {d\pi^{1/2}}\int_{-\infty}^{+\infty}dp_0
 n_F(\omega^0_\X +p_o/3,T)\int_{-\infty}^{0} dx
 e^{-\frac {x^2}{d^2}}\left(e^{6\g x}-1\right)
\label{ximxiw}\eea
where  the change variables to $x=-z_0+t_0/3$  was performed, and eq.
(\ref{Ndeux}) was used.
Expanding to first order in $3\g d$, it follows that,

\be \Xi- \Xi^{wall}\simeq \frac {3\g d} {\pi^{1/2}}\, \Xi\ll \Xi\qquad
\mbox{if}\qquad 3\g d\ll 1
\label{belle}\ee
where  eq. (\ref{densit}) was used.
The difference between $\Xi$ and $\Xi^{wall}$ consists in gaussian tails of
size $\sim d$,
 small compared to the range of damping $1/3\g$. The same phenomenon is present
 in $\Psi_{ref}$, through $\Psi_{inc}$, which appears in the convolution
(\ref{convol}).

\item Integrals on $z_0$ and $t_0>0$ for the reflected wave packet.

{}From eqs. (\ref{convol}), (\ref{wps}) and (\ref{borne}), it follows

\be \vert \Psi_{ref}(0,0;z_0,t_0,p_0,d,\X)\vert \le \frac m 2 \sqrt{\frac 1
{d\pi^{1/2}}}\int_0^{\infty} dz_1
e^{-\frac{(z_1-z_0+t_0/3)^2}{2d^2} +\g t_0}\ee

The contribution to the reflection probability, for $t_0>0$, is then bounded by

$$ n^\X_r(0,0;t_0>0)\le \frac {m^2} {2d
\Xi\pi^{1/2}}\int_{-\infty}^{+\infty} dp_0
2\g n_F(\omega^0_\X + p_0/3)\times $$\be
\int_0^{\infty} dz_1 dz_2 dz_0 dt_0
e^{-\frac{(z_1+z_0+t_0/3)^2+(z_2+z_0+t_0/3)^2}{2 d^2}+2\g
t_0}\label{doublexp}
\ee
where a change of variable $z_0\ra -z_0$ was performed. With the supplementary
variables changes
\be
z_1+z_0+t_0/3 = y d \cos\theta,\,\, z_2+z_0+t_0/3 = y d \sin\theta,
\,\,z_0+t_0/3=x d, \,\, t_0=t_0,\ee
the integral over $t_0$ leads to
\be  \int_0^{3xd}dt_0 e^{2 \g t_0} = \frac 1 {2\g}\left(e^{6\g x d}-1\right)
\simeq 3 x d ,\ee
while the integral on $x$,  assuming $\sin\theta\le\cos\theta$\footnote{ By
symmetry, the alternative sector, $\sin\theta\ge\cos\theta$ gives the same
contribution, leading to a factor of 2 in the final result.}, leads to
\be \int_0^{y \sin\theta} dx 3 x d^2 = \frac {3 d^2}  2 (y \sin\theta)^2 \ee
and, finally,
\be 3 d^4\int_0^{\infty}  dy y^3 e^{-\frac {y^2}2}\int_0^{\frac \pi 4}d\theta
\sin^2\theta= 3 d^4\frac {\pi-2} 4.\ee

\end{itemize}

It follows that

\be n^\X_r(z=0,t=0;t_0>0)\le \frac {m^2 d^3 6 \g }
{\pi^{1/2}}\left(\frac{\pi-2}8\right).\label{bornef}\ee

 When this result is compared to eq.(\ref{nrsimpl}),
\be n_r^\X(0,0)\simeq \frac {m^2} {\g \Xi}\ee
it is clear that the $t>0$ contribution is indeed suppressed,

\be \frac {n^\X_r(0,0;t_0>0)}{n_r^\X(0,0)}\sim  (\g d)^2 (\Xi d) \ll
1\label{final}\ee

where we have used the fact that $\Xi = O(g_sT)$ (a one-dimension density of
quasiparticles with energy in the $O(g_sT)$ range), and that it is  easy to
choose $d$ such that the condition (\ref{cond}) is met simultaneously with $(\g
d)^2 (\Xi d) \ll 1$. For example, $1/d\sim g_s T$ satisfies all these criteria.

\end{document}